\documentclass[preprint,12pt]{aastex}
\usepackage{natbib,epsfig}
\usepackage{graphicx}
\usepackage{multirow}
\tighten

\newcommand{\beq}{\begin{equation}}
\newcommand{\eeq}{\end{equation}}
\newcommand{\bea}{\begin{eqnarray}}
\newcommand{\eea}{\end{eqnarray}}
\newcommand{\g}{\gamma}
\newcommand{\gs}{\gamma_{self}}
\newcommand{\gt}{\widetilde{\gamma}}
\newcommand{\gmt}{\widetilde{\gamma_m}}

\newcommand{\gst}{\widetilde{\gamma_{self}}}

\newcommand{\nut}{\widetilde{\nu}}
\newcommand{\gh}{\widehat{\gamma}}
\newcommand{\gmh}{\widehat{\gamma_m}}
\newcommand{\goh}{\widehat{\gamma_0}}
\newcommand{\gch}{\widehat{\gamma_c}}
\newcommand{\gsh}{\widehat{\gamma_{self}}}
\newcommand{\numh}{\widehat{\nu_m}}
\newcommand{\nuch}{\widehat{\nu_c}}
\newcommand{\nuoh}{\widehat{\nu_0}}
\newcommand{\nuh}{\widehat{\nu}}
\newcommand{\epse}{\epsilon_e}
\newcommand{\epsB}{\epsilon_B}
\newcommand{\reps}{\frac{\epsilon_e}{\epsilon_B}}
\newcommand{\mrow}{\multirow}
\newcommand{\mcol}{\multicolumn}
\newcommand{\cent}{\centering}

\usepackage{ulem}
\usepackage{colordvi}

\shorttitle{Klein-Nishina effects on Synchrotron and Synchrotron
self-Compton spectrum} \shortauthors{Nakar, Ando \& Sari}

\begin{document}

\title{Klein-Nishina Effects on Optically Thin Synchrotron and Synchrotron
Self-Compton Spectrum}
\author{Ehud Nakar$^{1,2}$, Shin'ichiro Ando$^2$ and Re'em Sari$^{2,3}$}
\affil{1. Raymond and Beverly Sackler School of Physics \&
Astronomy, Tel Aviv University, Tel Aviv 69978, Israel\\
2. California
Institute of Technology, Mail Code 130-33, Pasadena, CA 91125, USA\\
3. Racah Institute for Physics, The Hebrew University, Jerusalem
91904, Israel\\}

\begin{abstract}
We present analytic approximations to the optically thin synchrotron
and synchrotron self-Compton (SSC) spectra when Klein-Nishina (KN)
effects are important and pair production and external radiation
fields can be neglected.  This theory is useful for analytical
treatment of radiation from astrophysical sources, such as gamma-ray
bursts (GRBs), active galactic nuclei and pulsar wind nebula, where
KN effects may be important. We consider a source with a continuous
injection of relativistic electrons with a power-law energy
distribution above some typical injection energy. We find that the
synchrotron-SSC spectra can be described by a broken power-law, and
provide analytic estimates for the break frequencies and power-law
indices. In general, we show that the dependence of the KN
cross-section on the energy of the upscattering electron results in
a hardening of the energy distribution of fast cooling electrons and
therefore in a hardening of the observed synchrotron spectrum. As a
result the synchrotron spectrum of fast cooling electrons, below the
typical injection energy, can be as hard as $F_\nu \propto \nu^0$,
instead of the classical $\nu^{-1/2}$ when KN effects are neglected.
The synchrotron energy output can be dominated by electrons with
energy above the typical injection energy. We solve
self-consistently for the cooling frequency and find that the
transition between synchrotron and SSC cooling can result in a
discontinuous variations of the cooling frequency and the
synchrotron and SSC spectra. We demonstrate the application of our
results to theory by applying them to prompt and afterglow emission
models of GRBs.

\end{abstract}

%\keywords{gamma-rays: bursts --- radiation mechanisms: non-thermal}

%%%%%%%%%%%%%%%%%%%%%%%%%%%%%%%%%%%%%%%%%%%%%%%%%%%%%%%%%%%%%%%%%%%%
\section{Introduction}
Synchrotron and synchrotron self-Compton (SSC) are common radiation
processes in astrophysical environments where  relativistic
electrons are continuously injected into a magnetized plasma. In the
Thomson scattering regime, the SSC component dominates the energy
output whenever the local energy density of the synchrotron photons
is larger than the energy density of the local magnetic field, as
well as the energy density of any external radiation field. In such
a case, synchrotron photons that are upscattered once by the
synchrotron emitting electrons carry more energy than the
unscattered synchrotron photons. The typical frequency of the
upscattered photons is increased by a factor of $\sim \gamma^2$,
where $\gamma$ is the typical electron Lorentz factor. Photons that
are upscattered twice carry more energy than those that are
scattered once and there typical frequency is again multiplied by
$\sim \gamma^2$. This hierarchical spectral structure, where the
total photons energy increases with the number of scattering and the
frequency of each generation is shifted by $\gamma^2$, continues up
to the point where the energy of the upscattered photons get to the
Klein-Nishina (KN) limit. Namely, the individual photon energy, as
measured in the rest frame of the upscattering electron, becomes
comparable to the electron rest-mass energy, $m_ec^2$. At this point
there is a transition of the scattering cross-section from the
constant Thomson regime to the Klein-Nishina regime, where it is
inversely proportional to the photons frequency. In addition, the
photon energy gain at each scattering (frequency shift), which is
proportional to the pre-scattering photon energy in the Thomson
regime, becomes constant (roughly $\g m_ec^2$) above the KN limit.
The cross-section fall-off and the saturation of the energy transfer
at each scattering, terminate the hierarchical spectral structure
above the KN limit.

The direct effect of the  KN limit on the observed spectrum is the
suppression of high-energy upscattered photons. However an indirect
effect is present in case that SSC emission dominates the energy
output, and that at least some of the injected electrons have enough
time to cool. In such a case electrons with different energies
(Lorentz factors) are cooling on a different fraction of the
radiation field. The reason is that some of the photons that are
below the KN limit for less energetic electrons are above this limit
for more energetic electrons. As a result, the energy distribution
of the cooling electrons is modified by the KN limit, and so does
the synchrotron spectrum as well as the spectra of all the SSC
hierarchical branches. While this indirect effect is less dramatic
than the high-energy KN cutoff, it may be the only observed KN
effect in cases that the high-energy SSC branches are above the
detectors energy range, thereby providing a valuable information
about the physical properties of the source. Moreover, the KN
modified synchrotron spectrum may be significantly different than
the unmodified one, making it necessary to include KN effects even
in cases where just the synchrotron spectrum is analyzed.

In relativistic sources, such as active galactic nuclei (AGNs) and
gamma-ray bursts (GRBs), it is thought that optically thin
synchrotron and SSC emission is produced behind relativistic shocks,
where a fresh population of relativistic electrons is injected.
Analytic approximation of the synchrotron-SSC spectra in such
systems is rather complex even when the KN feedback on the electron
distribution are not fully accounted for
\citep[e.g.,][]{Meszaros94,Tavecchio98,Petry00,Dermer00,Sari01,Bottcher02,Fan06,Zhang07}.
\cite{Rees67} derived the set of equations which self-consistently
follow the electron distribution for a one zone synchrotron-SSC
model (when pair-production can be ignored) and solved them
numerically for several examples in the context of radio sources.
Numerical calculations for different sources (e.g., GeV AGNs and
GRBs), and with additional physics (e.g., time dependent photon
field and pair production) followed
\cite[e.g.,][]{Coppi92,Mastichiadis97,Li00,Peer05a,Fan08,Vurm08}.
However, useful analytic approximations of the optically thin
synchrotron-SSC spectra in cases that KN effects are important, were
only partially discussed for a few special
cases\citep{Derishev03,Peer05b,Ando08}.

The purpose of this paper is to provide an analytic approximation of
the optically thin synchrotron and SSC spectra in cases where KN
feedback plays an important role and cannot be ignored, while pairs
production can be ignored. We consider a relativistic blast wave
that injects electrons with a power-law energy distribution and
provide a comprehensive analytic approximate description of the
resulting spectra as a function of the importance of the KN
suppression and the relative cooling by synchrotron and SSC. We
provide an analytic formula for the SSC to synchrotron energy output
ratio as a function of the physical parameters. These approximations
are useful as guidelines of the expected range of possible optically
thin synchrotron-SSC spectra and as a tool for interpretation of
observation without carrying out elaborate numerical calculations.
The limitation of the analytic approach is that it provides a sharp
broken power-law spectra,  while the true spectra are quite smooth.
Since KN effects on the electron spectrum are resulting from
integration over the electron spectrum itself (see Eq. \ref{EQ
Ynumeric}), the synchrotron power-law-breaks that are associated
with KN effects are smoother than those that are not. Nevertheless,
the approximated analytic spectra provided here are by far more
accurate than calculations (numerical or analytic) that neglect KN
effects on the electron spectrum. We caution that these calculations
apply only when the source is optically thin and pair production can
be neglected. This is not a trivial demand, since photons that are
upscattered near the KN limit are energetic enough to produce pairs
with the seed photons. Thus, pair production can be neglected only
when the source is optically thin not only to Thomson scattering on
electrons/pairs but also to pair creation by energetic photons on
the seed photons. While the later condition is more stringent (since
seed photons are more numerous than optically thin electrons/pairs),
it is still applicable to many astrophysical sources.

Our results show that since lower-energy electrons are cooling more
efficiently by SSC emission, KN effects result in harder spectra of
electrons that are cooling fast (i.e., radiating most of their
energy over the system lifetime). We show that in case that
electrons are continuously injected with some minimal Lorentz factor
$\g_m$ and all the electrons are cooling fast, the spectrum of the
synchrotron flux at frequencies that are below $\nu_{syn}(\g_m)$ can
be, in extreme cases, as hard as $F_\nu \propto \nu^{0}$ (compared
to the typical $F_\nu \propto \nu^{-1/2}$, where $\nu_{syn}$ is the
synchrotron frequency). We also show that the spectrum at
$\nu>\nu_m$ can be hard enough so most of the synchrotron energy is
emitted at $\nu>\nu_m$ even if most of the injected electron energy
is in electrons with Lorentz factor of the order of $\g_m$ which are
cooling fast. Another KN effect is that the transition from
electrons that are cooling fast by synchrotron emission to electrons
that are cooling fast by SSC emission can result in a dramatic
observational signature where the observed spectral break that
corresponds to the cooling Lorentz factor (above which electrons are
cooling fast) ``jumps" by orders of magnitude within a short time.

KN effects increase the complexity of the observed spectra and the
number of different types of spectra. In the paper we separate the
observed spectra to six different types, which covers the most
relevant possibilities, each corresponding to different physical
conditions in the source. For convenience, table \ref{Table physical
cond} summarizes the observed spectral types that correspond each to
a given physical system and gives the references to the equations
that are relevant to each case. Tables \ref{Table Critical
gammas}-\ref{Table SSC PLSs} list the values of the break
frequencies and of the spectral power-law indices for each of the
different cases. A reader that is interested only in the specific
spectrum of a given physical system can start the search in table
\ref{Table physical cond}.

The paper is organized as follows: in \S\ref{SEC spectra guidelines}
we describe the physical model and present guidelines of the general
considerations that we use to derive the synchrotron-SSC analytic
spectra in the different cases. Spectra of systems where all the
electrons are cooling fast are derived in \S\ref{SEC fastcooling}
and spectra of systems where most of the energy is in electrons that
cool slowly are presented in \S\ref{SEC slowcooling}. The exact
definition and value of the cooling frequency, as well as its
effects on the observed spectrum, are discussed in \S\ref{SEC g_c}.
We demonstrate the application of our results to theory by applying
them to prompt and afterglow emission models of GRBs in \S\ref{SEC
GRBs}. The main results are summarized in \S\ref{SEC summary}.

\section{Synchrotron-SSC spectra --- general considerations}\label{SEC spectra guidelines}

We consider a one-zone model where relativistic electrons radiate
synchrotron emission and where the inverse-Compton emission of these
electrons is dominated by upscattering their own emitted synchrotron
photons. The electron population in the source is generated by a
continuous injection of relativistic electrons with a power-law
distribution, i.e., $Q \propto \gamma^{-p}$ for $\gamma>\gamma_m$
($Q=0$ otherwise) where $Q$ is the electron source function,
$\gamma$ is the electron Lorentz factor and $p>2$. We consider only
radiative cooling and define $\gamma_c$ as the Lorentz factor above
which electrons are cooled efficiently over the age of the
system.\footnote{As a result of KN effects the definition of
$\gamma_c$ may be more complicated in a small region of the
parameter space. We discuss the definition (and value) of $\gamma_c$
in \S\ref{SEC g_c}.} The source is optically thin to photons emitted
at the typical synchrotron frequency so self-absorbtion can be
ignored. The source is also optically thin for pair production by an
arbitrarily energetic photon on the synchrotron photons so pair
production can be neglected. We also neglect short time scale
variability of the photon field in the sense that we work in the
regime where the photon field is determined by the instantaneous
electron distribution. Finally, cosmological redshift effects are
neglected in this study. These can be easily added to the results
according to the usual prescription.

The scenario of interest here is the one where the energy density of
synchrotron photons in the source is larger than the magnetic field
energy density, since only in this regime the electrons distribution
depends on KN effects as the cooling of some of the electrons is
dominated by SSC while others are cooled predominantly by
synchrotron radiation. Therefore throughout the paper we assume the
synchrotron energy density dominates over the magnetic field energy
density, with the exception of \S\ref{SEC dominantsynch fast} and
\S\ref{SEC dominantsynch slow} where, for completeness, we briefly
discuss the SSC spectra when synchrotron radiation dominates the
cooling of all electrons. Finally we discuss here in detail only the
case where the KN limit is relevant for the synchrotron photons,
implying that multiple scattering  are strongly suppressed and only
single scattering SSC is important. The case where there are
multiple SSC generations is not discussed here but it shows similar
effects and similar values of the asymptotic power-law indices to
those of the single generation case.

The self-consistent electron distribution and the observed spectra
can be easily computed once we know the synchrotron to SSC
emissivity ratio as a function of the electron Lorentz factor:
\begin{equation}\label{EQ Ydef}
   Y(\gamma) \equiv \frac{P_{SSC}(\gamma)}{P_{syn}(\gamma)}.
\end{equation}
When KN effects are neglected $Y \equiv Y_{noKN}$ is constant and we
first rederive its value \cite[e.g.,][]{Sari01}. The scenario that
we have in mind is an emission from a plasma that is heated by a
relativistic shock where relativistic electrons are injected at the
shock front and carry a fraction $\epsilon_e$ out of the injected
internal energy. The fraction of the magnetic field energy out of
the internal energy in the shocked plasma is $\epsilon_B$. In this
case
\begin{equation}\label{EQ YnoKN}
   Y_{noKN} \approx \frac{U_{syn}}{U_B}\approx\frac{\eta \epsilon_e}{(1+Y_{noKN})\epsilon_B}  .
\end{equation}
where $U_{syn}$ and ${U_B}$ are the energy density of synchrotron
photons and magnetic field respectively and
$\eta=\min\{1,(\gamma_m/\gamma_c)^{p-2}\}$ is the fraction of
electrons energy that was emitted. Therefore:
\begin{equation}\label{EQ YnoKN2}
   Y_{noKN} \approx \left\{\begin{array}{cc}
                      \sqrt{\frac{\eta \epsilon_e}{\epsilon_B}}& Y_{noKN} \gg 1\\
                      \frac{\eta \epsilon_e}{\epsilon_B}& Y_{noKN} \ll 1
                    \end{array}\right. .
\end{equation}
When relating $\frac{U_{syn}}{U_B}$ to
$\frac{\epsilon_e}{\epsilon_B}$ we assume that the volume occupied
by the shocked plasma, $V_{plasma}$, is comparable to the volumed
occupied by synchrotron radiation, $V_{rad}$. This is true for the
relativistic shocks considered here but is not necessarily correct
in the general case. In case that $V_{plasma} \ll V_{rad}$ then Eq.
\ref{EQ YnoKN} is generalized by multiplying the right hand side by
the fraction of the photon energy  that is contained within
$V_{plasma}$. In case of a non-relativistic blast wave with velocity
$v_{sh}$ this fraction is roughly $v_{sh}/c$ \citep{Sari01}. This is
the only generalization needed in order to apply the results of this
paper to cases where $V_{plasma} \ll V_{rad}$.

The line of reasoning used to derive Eq. \ref{EQ YnoKN} can be used
to show that in general when electrons with Lorentz factor $\gamma$
are fast cooling by upscattering their own synchrotron photons then
$Y(\g)(1+Y(\g)) \approx \epse(\g)/\epsB$ where $\epse(\g)$ is the
fraction of energy injected in electrons with Lorentz factor $\g$.

In order to understand the effects of the KN limit it is useful to
define the following function of an electron Lorentz factor:
\begin{equation}\label{EQ gamma_hat}
    \gh=\frac{m_ec^2\Gamma}{h\nu_{syn}(\gamma)}\propto \gamma^{-2},
\end{equation}
and its reciprocal function:
\begin{equation}\label{EQ gamma_tilde}
   \gt=\left(\frac{\gamma m_ec^2\Gamma}{h\nu_{syn}(\gamma)}\right)^{1/2}\propto \gamma^{-1/2},
\end{equation}
where $m_e$ is the electron mass, $c$ is the light speed, $h$ is
Planck's constant, $\Gamma$ is the bulk Lorentz factor of the source
(assuming the source is moving towards the observer) and
$\nu_{syn}(\gamma) \approx \Gamma\gamma^2 q_e B/(2\pi m_e c)$ is the
typical observed frequency of the synchrotron emission of an
electron with Lorentz factor $\gamma$ ($B$ is the magnetic field and
$q_e$ is the electron charge). These functions are useful since
synchrotron photons emitted by electrons with Lorentz factor larger
than $\gamma$ (i.e., with frequency $>\nu_{syn}(\g)$) cannot be
upscattered efficiently by electrons with Lorentz factor larger than
$\gh$ because they are above the KN limit. Similarly, electrons with
Lorentz factor $\gamma$ cannot efficiently upscatter synchrotron
photons emitted by electrons with Lorentz factor larger than $\gt$.

The KN limit introduces new critical Lorentz factors in addition to
$\g_m$ and $\g_c$:
\begin{equation}\label{EQ gamma_m_hat}
  \begin{array}{c}
    \gmh = \frac{m_ec^2\Gamma}{h\nu_m} \\
    \\
       \gch = \frac{m_ec^2\Gamma}{h\nu_c},
  \end{array}
\end{equation}
where $\nu_m \equiv \nu_{syn}(\g_m)$ and $\nu_c \equiv
\nu_{syn}(\g_c)$,
\begin{equation}\label{EQ gamma_self}
    \gs = \left(\frac{B_{QED}}{ B}\right)^{1/3}=\g^{2/3} \gh^{1/3}
\end{equation}
which satisfies $\gsh=\gst=\gs$ and $\g_0$ which satisfies
\begin{equation}\label{EQ gamma_0}
    Y(\g_0) = 1.
\end{equation}
$B_{QED}= 2\pi m_e^2 c^3/(q_e h)=4.4 \times 10^{13}$ G is the
quantum critical field. The requirement in this paper that the
energy density of synchrotron photons in the source is larger than
the magnetic field energy density guaranties that $\g_0$ is well
defined. Note that this requirement is equivalent to $Y_{noKN}>1$
and therefore to $\eta\epse/\epsB>1$.

When KN effects are important, $Y(\gamma)$ is not a constant anymore
and it affects the electron radiative cooling function:
\begin{equation}\label{EQ cool func}
    \frac{d\g}{dt} = -\frac{\sigma_T B^2}{6\pi m_e c}~\g^2\left[1+Y(\g)\right].
\end{equation}
where $t$ is the time as measured in the source rest frame
($t/2\Gamma$ is the time in the observer frame), $\sigma_T$ is the
Thomson cross section and $B$ is the magnetic field at time $t$. The
continuity equation of the electron distribution reads:
\begin{equation}\label{EQ dN_dt}
    \frac{\partial N_\g}{\partial t} + \frac{\partial
    }{\partial \g } \left(N_\g\frac{d\g}{dt}\right)= Q,
\end{equation}
where $N_\g$ is the electron number per unit of $\g$. Here we
consider
\begin{equation}\label{EQ Q}
    Q = Q_0 \left\{\begin{array}{cc}
               0 & \g<\g_m \\
               \left(\frac{\g}{\g_m}\right)^{-p} & \g>\g_m
              \end{array}\right. ,
\end{equation}
where $Q_0$ is constant and $p>2$. Solving Eqs. \ref{EQ
dN_dt}--\ref{EQ Q}, we find that the electron distribution can be
approximated in the fast cooling regime (i.e., $\g_c<\g_m$) as:
\begin{equation}\label{EQ Ng_fast}
    N_\g \propto
    \frac{1}{1 + Y(\gamma)}\left\{
       \begin{array}{lr}
         \gamma^{-2} & \g_c < \g < \g_m \\
         \gamma^{-p-1} & \g_m < \g
       \end{array}\right.
\end{equation}
while in the slow cooling regime (i.e., $\g_m<\g_c$) it is
\begin{equation}\label{EQ Ng_slow}
    N_\g \propto
    \left\{
       \begin{array}{lr}
         \gamma^{-p} & \g_m < \g < \g_c \\
         \frac{1}{1 + Y(\gamma)}\;\gamma^{-p-1} & \g_c < \g
       \end{array}\right.
\end{equation}
where in both regimes there are no electrons with
$\g<\min\{\g_c,\g_m\}$. Equations \ref{EQ Ng_fast} and \ref{EQ
Ng_slow} imply that the electron distribution depends significantly
on $Y(\g)$ only if both $\g_c<\g$ and $Y(\g)>1$ are satisfied (i.e.,
$\g_c<\g<\g_0$).

In general the $Y$ parameter  can be approximated, in case of an
isotropic photon field and ultra-relativistic electrons, as:
\begin{equation}\label{EQ Ynumeric_smooth}
   Y(\g) =
%   \frac{1}{U_B}\int_0^\infty \int_{-1}^1 \chi\left(\frac{\nu}{\nut},\mu\right) \frac{dU_{ph}}{d\nu d\mu} d\mu d\nu
    \frac{1}{U_B}\int_0^\infty \frac{dU_{ph}}{d\nu}
    \int_{-1}^1\frac{ \frac{1}{2 \sigma_T} (1-\mu) \sigma_{KN} \left[\frac{\nu}{\nut}(1-\mu)\right]
   }{1+\frac{\nu}{\nut}} d\mu d\nu
\end{equation}
where $\mu$ is the cosine of the angle between the upscattered
photon and the scattering electron momenta in the source frame and
$\sigma_{KN}[x]$ is the Klein-Nishina cross-section for scattering
of photons with energy $h\nu=x m_e c^2$ in the electron's rest
frame. The numerator in the integral over $\mu$ gives the rate at
which photons with a frequency $\nu$ and incident angle $\mu$ are
upscattered, normalized to the rate in the Thomson scattering
regime. The factor $1/(1+\nu/\nut)$ is approximately the energy
given to each upscattered photon, again normalized to the energy
given in the Thomson regime, where in the Thomson regime ($\nu/\nut
<< 1$) photons gain energy that is proportional to their initial
frequency, while in the Klein-Nishina regime ($\nu/\nut > 1$)
photons gain roughly a constant amount of energy ($\gamma m_e c^2$).
Therefore in the Thomson regime the integral over $\mu$ gives $1$
while at the Klein-Nishina regime it is $\propto 1/\nu^2$
(neglecting logarithmic terms; note that $\sigma_{KN}[x \gg 1]
\propto \ln(2x)/x$). Therefore if $d \ln(U_{ph})/d \ln (\nu) < 2$
for all $\nu>\nut$ then the integral over $\mu$ can be approximated
as a Heaviside step function $H(\nut-\nu)$ and Eq. \ref{EQ
Ynumeric_smooth} reads $Y(\g) \approx U_{ph}[\nu<\nut]/U_B$.

If in addition to the step function approximation we farther assume
that there is no short time scale variability of the photon field
(namely that the instantaneous emissivity determines the photon
density) and that the volume occupied by the shocked plasma is equal
to the volume occupied by the synchrotron radiation, we can write:
\begin{equation}\label{EQ Ynumeric}
   Y(\g) =
   \frac{U_{ph}[\nu<\nut]}{U_B}=
   \reps \frac{(p-2)\int_0^{\nut\,'(\g)} \int P_{\nu',syn}^{'}(\g^* ) N_{\g^*} d\g^* ~d\nu\,'}{Q_0m_ec^2\g_m^2},
\end{equation}
where $P_{\nu',syn}^{'}(\g^*)$ is the synchrotron emissivity per
frequency unit of an electron with Lorentz factor $\g^*$ and
$\nut\,'(\g)$ is the synchrotron frequency of $\gt$ electron (note
that both $\nut\,'$ and $\gt$ are functions of $\g$), both measured
in the plasma rest frame ($\nut\,$ is the corresponding frequency as
measured in the observer frame). The term $Q_0m_ec^2\g_m^2/(p-2)$ is
the total electron energy injected into the emitting region  per
unit of time. Since we are looking for approximated power-law
spectra the integrals of this equation can be approximated so:
\begin{equation}\label{EQ Y}
   Y(\g)\propto\left\{\begin{array}{cc}
            {\rm const} & \nut > \nu_{peak}^{syn} \\
            \g^{-\alpha(\gt)-1} & \nut<\nu_{peak}^{syn}
           \end{array}\right. ,
\end{equation}
where
\begin{equation}\label{EQ alpha}
  \alpha(\gt) = \left.
  \frac{{\rm d} \ln (F_\nu)}{{\rm d}
  \ln (\nu)}\right|_{\nut} ,
\end{equation}
$F_\nu$ is the observed energy flux per unit frequency and
$\nu_{peak}^{syn}$ is the observed frequency which dominates the
synchrotron energy output. Equation \ref{EQ Y} can be understood as
follow. As long as $\nut>\nu_{peak}^{syn}$ the value of
$U_{ph}[\nu<\nut]$ is dominated by $\nu_{peak}^{syn}$ photons and is
therefore roughly constant. These electrons can inverse-Compton the
photons containing most of the energy without suffering from the KN
reduction. For $\nut<\nu_{peak}^{syn}$ the SSC emissivity of $\g$
electrons is dominated by upscattering $\nut$ photons, implying
$Y(\g)\propto \left. \nu F_\nu\right|_{\nut}$ and since $\nut
\propto \g^{-1}$ we obtain $Y(\g) \propto \g^{-\alpha(\gt)-1}$. As
discussed above, the approximation of the KN limit as a step
function is valid as long as\footnote{We use the step function
approximation also when we numerically integrate over Eq. \ref{EQ
Ynumeric}  in order to evaluate the spectrum. We tested serval cases
to confirm that taking the accurate KN cross-section and the average
over all photons incident angles does not significantly affect the
conclusions we draw based on numerical results. Its only effect in
the cases discussed in this paper is to produce smoother light
curves.} $\alpha(\gt) < 1$. For $\alpha(\gt) > 1$ the increase in
the photon energy density at frequencies larger than $\nut$
overcompensate for the KN reduction, so the emissivity of $\g$
electron is actually dominated by upscattering of photons with
$\nu>\nut$ deep in the KN regime and $Y(\gamma)\propto \gamma^{-2}$.
Since in the spectral regimes considered in this paper $\alpha(\gt)
< 1$ is always satisfied we use Eqs. \ref{EQ Ynumeric} and \ref{EQ
Y} throughout the paper. Self absorption results in spectra of
$\alpha=11/8$, $\alpha=2$ or $\alpha=5/2$ \cite[see
e.g.,][]{Granot00}, and therefore all result in $Y(\gamma) \propto
\gamma^{-2}$. We do not discuss self absorption any farther in this
paper. Finally in order to close the set of equations we need to
relate the power-law index of the synchrotron spectrum at
$\nu_{syn}(\g)$ to the electron distribution $N_\g$:
\begin{equation}\label{EQ Fnusyn_general}
    \alpha(\g)=\frac{1}{2}\left(\frac{{\rm d}\ln (N_\g)}{{\rm
    d}\ln(\g)}+1\right) .
\end{equation}

A self consistent solution of equations \ref{EQ Ng_fast} (or \ref{EQ
Ng_slow}), \ref{EQ Y} and \ref{EQ Fnusyn_general} can provide an
analytic approximation of the observed spectrum. Examining those
equations we can define several simple rules that will enable us to
find the critical Lorentz factors in each case, and the
corresponding frequencies where there are breaks in the synchrotron
spectrum. First, in addition to the usual breaks at (the frequencies
corresponding to) $\g_m$ and $\g_c$ there will be a break at $\g_0$
if $\g_0>\g_c$. Next, a break in the synchrotron spectrum at some
$\g_{b}$ has a corresponding break in $Y(\g)$ if $\nu_{syn}(\g_{b})
\leq \nu^{syn}_{peak}$ (Eq. \ref{EQ Y}). Now, a break in $Y(\g_b)$
results in a break in the electron distribution at
$\widehat{\g_{b}}$ in case that $\g_c<\widehat{\g_{b}}<\g_0$ (Eqs.
\ref{EQ Ng_fast} and \ref{EQ Ng_slow}). Thus, a break at some
$\g_{b}$ has a corresponding critical frequency at
$\widehat{\g_{b}}$ given $\g_c<\widehat{\g_{b}}<\g_0$ and
$\nu_{syn}(\g_{b}) \leq \nu^{syn}_{peak}$. In principle, there can
be a series of critical frequencies at $\g_{b},~\widehat{\g_{b}},
~\widehat{\widehat{\g_{b}}},~...$ . The fact that $\gh \propto
\g^{-2}$ ensures that the series is terminated at some point with a
frequency that is larger than $\g_0$ or smaller than $\g_c$.
Furthermore, critical frequencies of second order or higher (i.e.,
$\widehat{\widehat{\g_{b}}}$,
$\widehat{\widehat{\widehat{\g_{b}}}}$, etc.) can usually be
neglected since they correspond to very mild breaks. Based on this
algorithm to find the critical frequencies we can see that there are
different types of spectra that are determined by the relations
between $\g_m$, $\g_c$, $\g_0$, $\gmh$, $\gch$ and $\goh$. As it
turns out it is enough to know $\g_m$, $\g_c$, $\reps$, and one
additional KN frequency, e.g, $\g_{self}$, $\gmh$ or $\gch$, to
determine the relation between all different frequencies and to
describe the entire spectral shape. While physically it is more
natural to use $\g_{self}$ we use $\gmh$ and/or $\gch$ since there
are  observable spectral features corresponding to these Lorentz
factors.

Below we go over all possible relations between the critical
frequencies and find six general types of synchrotron spectra. We
shall discuss separately slow ($\g_m<\g_c$) and fast ($\g_c<\g_m$)
cooling regimes, where the fast cooling regime is separated to cases
where $\g_m<\gmh$, $\g_m=\gmh$ and $\g_m>\gmh$. The latter is the
most complicated case and we divide it further into three subcases.
For each case we first present the value of $Y(\gamma)$, then the
synchrotron spectrum and finally the SSC spectra. The value of
$\gamma_c$ and its evolution for each of the cases are discussed
separately in \S\ref{SEC g_c}. For convenience we use the notation
$\nu_x$ to denote the synchrotron frequency that corresponds to an
electron with Lorentz factor $\g_x$. For example
$\nu_0\equiv\nu_{syn}(\g_0)$, $\numh\equiv\nu_{syn}(\gmh)$,
$\nuch\equiv\nu_{syn}(\gch)$ etc.

\section{Fast cooling spectra ($\g_c<\g_m$)}\label{SEC fastcooling}

\subsection{Case I - weak KN regime: $\gamma_m < \gmh$}\label{SEC caseI}

Here  $\g_0$ and $\gch$ are always larger than  $\gmh$ but the order
between them may vary. $\widehat{\gch}<\g_c$ and is therefore
unimportant. Typically $\nu_{peak}^{syn} = \nu_m$ (unless $p \approx
2$ and $\gmh/\g_m$ is not too large; we discuss this special case in
the appendix) in which case $\goh$ and $\widehat{\gmh}$ are larger
than $\nu_{peak}^{syn}$ and are therefore irrelevant. $\g_m$
electrons are cooling primarily on $\nu_m$ photons implying $Y(\g_m)
\approx (\epsilon_e/\epsilon_B)^{1/2}$. When $\nu_{peak}^{syn} =
\nu_m$ then all electrons with $\g<\gmh$ are also cooling on $\nu_m$
photons implying a constant $Y(\g<\gmh)$. There are breaks in $Y$ at
$\gmh$ and $\gch$ that corresponds to spectral slopes
$\alpha(\g_c<\g<\g_m)=-1/2$ and $\alpha(\g<\g_c)=1/3$ respectively:
\begin{equation}\label{EQ Y1}
    Y(\gamma)=\left\{\begin{array}{lc}
                       \sqrt{\frac{\epsilon_e}{\epsilon_B}} & \g< \gmh \\
                       \sqrt{\frac{\epsilon_e}{\epsilon_B}} \left(\frac{\g}{\gmh}\right)^{-1/2} &
                       \gmh<\gamma<\gch\\
                       \sqrt{\frac{\epsilon_e}{\epsilon_B}}\frac{\g_c}{\g_m}\left(\frac{\g}{\gch}\right)^{-4/3} &  \gch<\gamma
                    \end{array}\right. .
\end{equation}
According to this $Y$ distribution:
\begin{equation}\label{EQ g0I}
   \g_0 = \left\{\begin{array}{lc}
                   \gmh\frac{\epse}{\epsB} & \sqrt{\frac{\epse}{\epsB}}<\frac{\g_m}{\g_c} ~;~(\g_0<\gch) \\
                   \gch\left(\frac{\g_c}{\g_m}\right)^{3/4}\left(\frac{\epse}{\epsB}\right)^{3/8} & \sqrt{\frac{\epse}{\epsB}}>\frac{\g_m}{\g_c} ~;~(\g_0>\gch)
                 \end{array}\right.
\end{equation}

\begin{figure*}
\includegraphics[width=16.5cm]{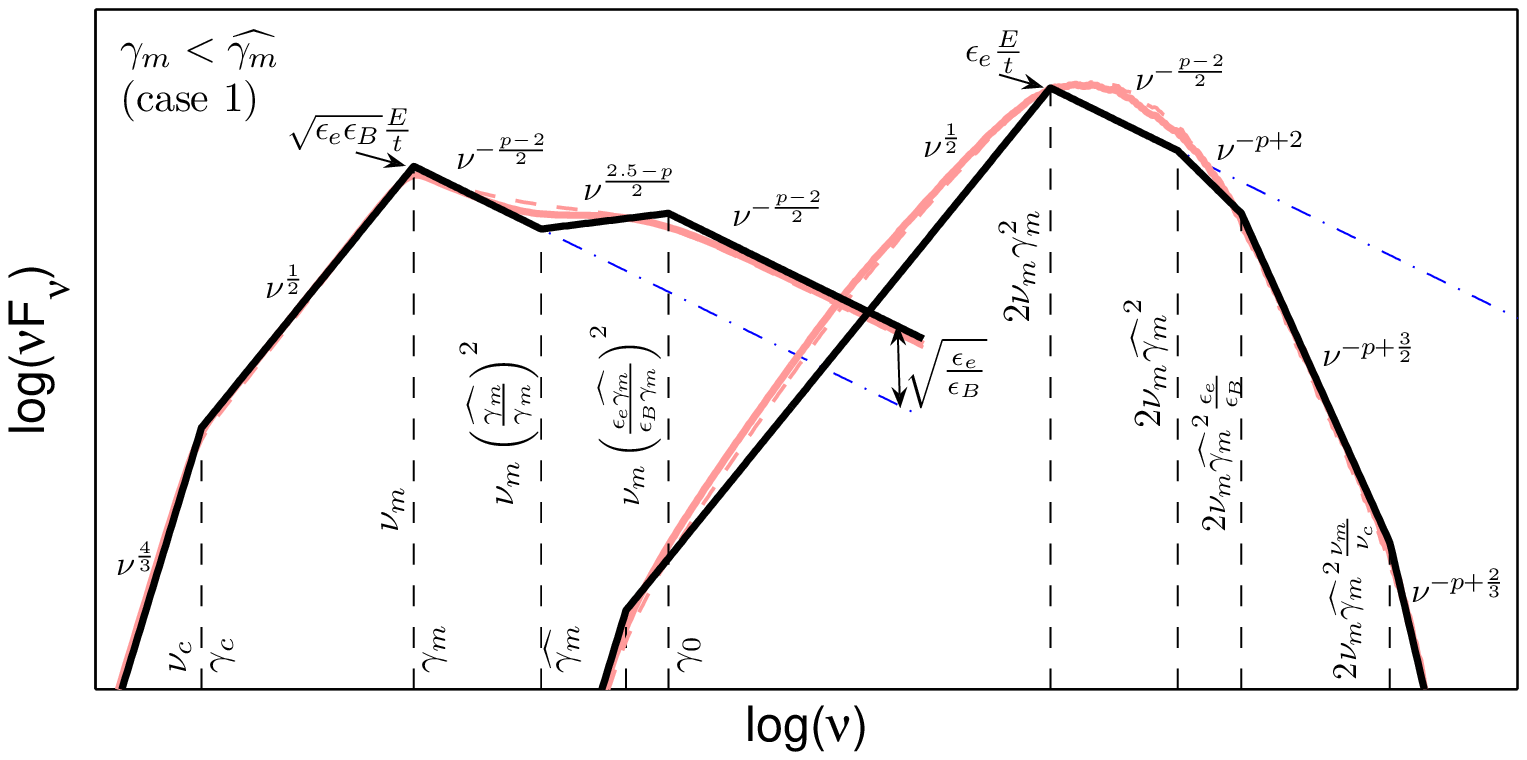}
  \caption{The synchrotron-SSC spectrum for $\gamma_c < \gamma_m <
  \gmh$ (Case I) and $\g_0<\gch$. The specific parameters are
  $\g_m/\gmh=10^{-3}$, $\epse/\epsB=10^{3}$ and $p=2.4$.
  The analytic spectrum ({\it black line}; see text) is compared
  to the numerical spectra, calculated by numerically integrating Eqs.
  \ref{EQ cool func}, \ref{EQ dN_dt} and \ref{EQ Ynumeric} ({\it solid red line})and by integrating over
  \ref{EQ cool func}, \ref{EQ dN_dt} and \ref{EQ Ynumeric_smooth} ({\it dashed red
  line}). It is evident that the numerical spectra are smooth, where taking the actual KN cross-section
  into account (Eq. \ref{EQ Ynumeric_smooth}) results in an
  additional smoothing over the step function approximation (Eq. \ref{EQ Ynumeric}). The
  value  of the critical synchrotron frequencies is written next to the value of the
  corresponding electron Lorentz factor. The dash-dot lines are the
  spectrum in case that KN effects are ignored. Also noted are the synchrotron luminosity
  ($\sqrt{\epse/\epsB}E/t$), SSC luminosity ($\epse E/t$) and the ratio of the
   synchrotron spectrum at $\nu_0\ll\nu$ with and without including the
   KN limit ($\sqrt{\epse/\epsB}$) .}\label{FIG case1}
\end{figure*}

The corresponding synchrotron spectrum always has spectral breaks at
$\nu_c$, $\nu_m$, $\nuh$ and $\nu_0$. In case that $\g_0<\gch$
(i.e., $(\epse/\epsB)^{1/2}<\g_m/\g_c$) these are the only break
frequencies and the synchrotron spectrum is
\begin{equation}\label{EQ Fsyn1a}
    F_{\nu}^{syn} \propto \left\{\begin{array}{lc}

      \nu^{-1/2} & \nu_c<\nu<\nu_m\\
      \nu^{-p/2} &  \nu_m<\nu<\numh\\
      \nu^{-(p/2-1/4)} & \numh<\nu<\nu_0\\
      \nu^{-p/2} &  \nu_0<\nu
    \end{array}\right.
\end{equation}
An example of the analytic synchrotron spectrum, based on Eq.
\ref{EQ Fsyn1a} (with $\g_0<\gch$), and a comparison to numerically
calculated spectrum is presented in figure \ref{FIG case1}.

An additional break at $\nuch$ is observed in case that $\gch<\g_0$
(i.e., $(\epse/\epsB)^{1/2}<\g_m/\g_c$):
\begin{equation}\label{EQ Fsyn1b}
    F_{\nu}^{syn} \propto \left\{\begin{array}{lc}

      \nu^{-1/2} & \nu_c<\nu<\nu_m\\
      \nu^{-p/2} &  \nu_m<\nu<\numh\\
      \nu^{-(p/2-1/4)} & \numh<\nu<\nuch\\
      \nu^{-(p/2-2/3)} & \nuch<\nu<\nu_0\\
      \nu^{-p/2} &  \nu_0<\nu
    \end{array}\right.
\end{equation}

{\it SSC spectrum}: As $\g_c,\g_m \ll \gmh$ the SSC luminosity in
this case is not significantly affected by KN cross-section. Most of
the energy is emitted around $2\nu_m \g_m^2$ and the spectrum is
similar to the one described in \cite{Sari01} up to
$\nu_{IC}(\gmh)=2\gmh^2\nu_m$. Electrons with $\gmh<\g<\g_0$ are
still radiating practically all their energy to SSC and therefore
the SSC spectrum is affected only mildly at $\nu \sim 2\gmh^2\nu_m$. The mild
SSC break at this frequency arises from the fact that electrons with
$\g <\gmh$ lose their energy to upscattering of $\nu_m$ photons, so
$\nu_{IC}(\g<\gmh)\propto \g^2$, while electrons with $\g>\gmh$ lose
their energy by upscattering $\nu_m \frac{\gmh}{\g}$ photons, so
$\nu_{IC}(\g>\gmh)\propto \g$. As a result the SSC spectral index at
$\nu>2\gmh^2\nu_m$ is $p-1$ compared to $p/2$ at lower frequencies.
If $\g_0<\gch$ then there is first a break at $\nu_{IC}(\g_0)$ to a
spectral index of $p-\frac{1}{2}$ and later at  $\nu_{IC}(\gch)$ to
a spectral index of $p+\frac{1}{3}$. Therefore the SSC spectrum for
$\g_0<\gch$ is:
\begin{equation}\label{EQ FIC1a}
    F_{\nu}^{IC} \propto \left\{\begin{array}{lc}
      \nu^{-p/2} & 2\nu_m\gamma_m^2<\nu<2\nu_m\gmh^2\\
      \nu^{-(p-1)} &  2\nu_m\gmh^2<\nu<2\nu_m \gmh \g_0\\
      \nu^{-(p-\frac{1}{2})} & 2\nu_m \gmh \g_0<\nu<2\nu_c\gch^2\\
      \nu^{-(p+\frac{1}{3})} &  2\nu_c\gch^2<\nu
    \end{array}\right.
\end{equation}
If $\gch<\g_0$ then there is no break at $\nu_{IC}(\gch)$ as the
faster increase in the number of electrons above $\gch$ compensates
for the rapid decrease in the flux below $\g_c$. There is a break at
$\nu_{IC}(\g_0)$ directly to an index of $p+\frac{1}{3}$ and the SSC
spectrum for $\g_c<\g_0$ is:
\begin{equation}\label{EQ FIC1b}
    F_{\nu}^{IC} \propto \left\{\begin{array}{lc}
      \nu^{-p/2} & \nu_m\gamma_m^2<\nu<\nu_m\gmh^2\\
      \nu^{-(p-1)} &  \nu_m\gmh^2<\nu<\nu_m \gmh \g_0\\
      \nu^{-(p+\frac{1}{3})} &  \nu_m \gmh \g_0<\nu
    \end{array}\right.
\end{equation}
An example of the analytic SSC spectrum when $\g_0<\gch$ (Eq.
\ref{EQ FIC1a}), and its comparison to numerically calculated
spectrum is presented in figure \ref{FIG case1}. The normalization
of the analytic spectrum is calculated using Eq. \ref{EQ
Ygm_complex}.

\subsubsection{Case I with $p\approx 2$}\label{SEC caseIp=2}
The discussion above is valid as long as $\nu_{peak}^{syn} = \nu_m$,
which is the more common case. Nevertheless, if
$(\g_m/\g_0)^{p-2}\sqrt{\epse/\epsB} > 1$ the synchrotron energy
output peaks at $\nu_0$ and additional power-law segments are
introduced. The exact spectrum depends on the ratio of energy
injected in electrons with Lorentz factor of order $\g_m$ to the
energy injected in electrons with Lorentz factor of order $\g_0$.
Here we present the spectrum in case that this ratio is $\approx 1$
(namely $(\g_m/\g_0)^{p-2}\approx 1$). In such case $\widehat{\gmh}$
and $\goh$ (both smaller than $\g_m$) become critical frequencies
(assuming that they are larger than $\g_c$). Electrons with
$\g<\widehat{\gmh}$ experience enhanced SSC cooling, thereby
suppressing the synchrotron emission at these frequencies. The
energy flux at $\nu_0$ is higher by a factor $\approx
(\epsilon_e/\epsilon_B)^{1/2}$ than the one at $\nu_m$, and
therefore the flux at $\nu<\nuoh$ is suppressed by the same factor.
As a result additional power-law segments are introduced to case I
spectra (we assume here $\g_0<\gch$):
\begin{equation}
\begin{array}{lc}\label{EQ YF1p=2}
  Y(\gamma)=\left\{\begin{array}{c}
                     \frac{\epsilon_e}{\epsilon_B} \\
                     \sqrt{\frac{\epsilon_e}{\epsilon_B}}\left(\frac{\g_m^2}{\gmh}\right)^{1/4}\g^{-1/4}
                   \end{array} \right. &
                   \begin{array}{c}
                     \gamma< \frac{\g_m^2}{\gmh}\left(\frac{\epsilon_e}{\epsilon_B}\right)^{-2} \\
                     \frac{\g_m^2}{\gmh}\left(\frac{\epsilon_e}{\epsilon_B}\right)^{-2}<\g< \frac{\g_m^2}{\gmh}
                   \end{array}\\
  F_\nu^{syn} \propto \nu^{-3/8} & \max\{\nu_c,\nuoh\}<\nu<\widehat{\numh}
\end{array} ~~~;~~~p\approx 2
\end{equation}
If $\gch<\g_0$ then anther power-law segment is introduced to $Y$,
but it does not affect the synchrotron spectrum.

Additional important property of this case is that the SSC to
synchrotron luminosity ratio is significantly reduced (and is
approximately unity), since electrons that are cooling primarily by
synchrotron emit comparable amount of energy to those that are
cooling by SSC (see discussion in \S\ref{SEC YbarFast}).

%Finally if $\widehat{\widehat{\gmh}}<\g_0$ it will become a critical frequency
%but as the resulting break is minor ($\alpha changes by no more than $1/16$) it can be ignored.

\subsection{Case II - strong KN regime: $\gmh < \g_m$}

In this regime $\gmh < \g_{self}<\g_m$ and $\gmh<\g_0$. The shape of
the spectrum depends mostly on the relations between $\g_0$,
$\g_{self}$ and $\g_m$. Therefore we divide this case to three
subcases: IIa) $\g_0 < \g_{self}<\g_m$, IIb) $\g_{self}<\g_0<\g_m$
and IIc) $\g_{self}<\g_m<\g_0$. The relevant case for a given set of
physical parameters is determined by the relation between the two
ratios $\g_m/\gmh$ and $\epse/\epsB$. The shape of the spectrum also
depends on the value of $\g_c$ relatively to the other critical
frequencies. Moreover, the time evolution of $\g_c$ depends on its
relative value. In this section we assume that in each of the
subcases $\g_c$ is small and has no significant effect on the
spectrum above $\nu_c$. The effects of $\g_c$ on the spectra in the
different cases and its evolution are discussed in \S\ref{SEC g_c}.

The SSC spectrum in this regime includes too meany subcases and
power-law segments to list them all in a useful way. However the
differences between the power-law indices  of the various segments
is typically small ($\leq 1/4$). Therefore for each subcase below we
give a rough description of the SSC spectrum, mostly near
$\nu_{peak}^{IC}$, which is accurate enough for comparison with
observation. A common feature of the SSC spectra in this regime
($\g_c,\gmh < \g_m$) is that most of the SSC energy is radiated by
$\g_m$ electrons at $\nu_{peak}^{IC} \approx 2\nu_m \g_m \gmh$.

Before discussing the specific subcases it is useful to note that in
this regime $\g_m$ electrons do not upscatter their own synchrotron
photons and therefore we cannot easily determine $Y(\g_m)$. However,
since in this regime $\nu_{self}<\nu_{peak}^{syn}$ electrons with
Lorentz factor $\g_{self}$ are primarily cooling on their own
emitted synchrotron photons (assuming $\g_c<\g_{self}$). The total
luminosity emitted by electrons with Lorentz factor of order $\g$ is
proportional to $\g N_\g(d\g/dt)$, and is therefore independent of
$Y$ and proportional to $\g$ for $\g_c<\g<\g_m$ (see Eqs. \ref{EQ
cool func} and \ref{EQ Ng_fast}). Therefore, based on the discussion
below Eq. \ref{EQ YnoKN2}, we can determine the value of
$Y(\g_{self})$:
\begin{equation}\label{EQ Y_gamma_KNself}
    Y(\g_{self})\left[1+Y(\g_{self})\right]\approx
    \frac{\epse}{\epsB}\frac{\g_{self}}{\g_m}=
    \frac{\epse}{\epsB}\left(\frac{\gmh}{\g_m}\right)^{1/3} .
\end{equation}

\subsubsection{Case IIa: $\gamma_0<\g_{self}<\g_m$
$~~\left[\frac{\epse}{\epsB}<\left(\frac{\g_m}{\gmh}\right)^{1/3}\right]$}\label{SEC caseIIa}

Since $\g_0$ is smaller than $\goh$, $\widehat{\gmh}$ and $\gch$,
synchrotron breaks (in addition to $\nu_m$ and $\nu_c$) correspond
only to $\gmh$ and $\g_0$. Electrons with $\g>\g_0$ cool by
synchrotron emission implying $F_\nu(\nu_0<\nu<\nu_m)\propto
\nu^{-1/2}$ and  $\alpha(\gt)=-1/2$ for electrons with
$\gmh<\g<\goh$. It follows that $Y(\gmh<\g<\goh) \propto \g^{-1/2}$
and from Eqs. \ref{EQ Ng_fast} and \ref{EQ Fnusyn_general},
$F_\nu(\numh<\nu<\nu_0) \propto \nu^{-1/4}$. As
$\gmh<\g_0,\g_{self}<\goh$ we can use Eq. \ref{EQ Y_gamma_KNself} to
find:
\begin{equation}\label{EQ g0IIa}
   \g_0 \approx \gmh\left(\reps\right)^2
\end{equation}
The modified synchrotron spectrum implies $Y(\goh<\g<\widehat{\gmh})
\propto \g^{-3/4}$ (note that $Y<1$ in this range). Finally the
value of $Y(\g<\gmh)=\epse/\epsB$ is constant since these electrons
are cooling on $\nu_m$ photons which are emitted by $\g_m$ electrons
that are predominantly cooling by synchrotron radiation. The
complete spectrum is therefore:
\begin{equation}\label{EQ Y2a}
    Y(\gamma) \approx \left\{\begin{array}{lc}
       \reps & \g< \gmh \\

       \reps \left(\frac{\g}{\gmh}\right)^{-1/2} &
       \gmh<\g<\frac{\g_m^2}{\gmh}\left(\reps\right)^{-4}\\

       \gmh^{1/4}\g_m^{1/2}\g^{-3/4} &
       \frac{\g_m^2}{\gmh}\left(\reps\right)^{-4}<\g<\frac{\g_m^2}{\gmh}\\

       \gmh^{1/2}\g^{-1/2} & \frac{\g_m^2}{\gmh}<\g\\
       \end{array}\right.
\end{equation}
and the synchrotron spectrum is modified only between $\numh$ and
$\nu_0$:
\begin{equation}\label{EQ Fsyn2a}
    F_\nu^{syn} \propto \left\{\begin{array}{lc}
                       \nu^{-1/2} & \nu_c<\nu<\numh \\
                       \nu^{-1/4} & \numh <\nu<\nu_0\\
                       \nu^{-1/2} & \nu_0<\nu<\nu_m\\
                       \nu^{-p/2} & \nu_m<\nu
                       \end{array}\right.
\end{equation}

\begin{figure}
\includegraphics[width=16.5cm]{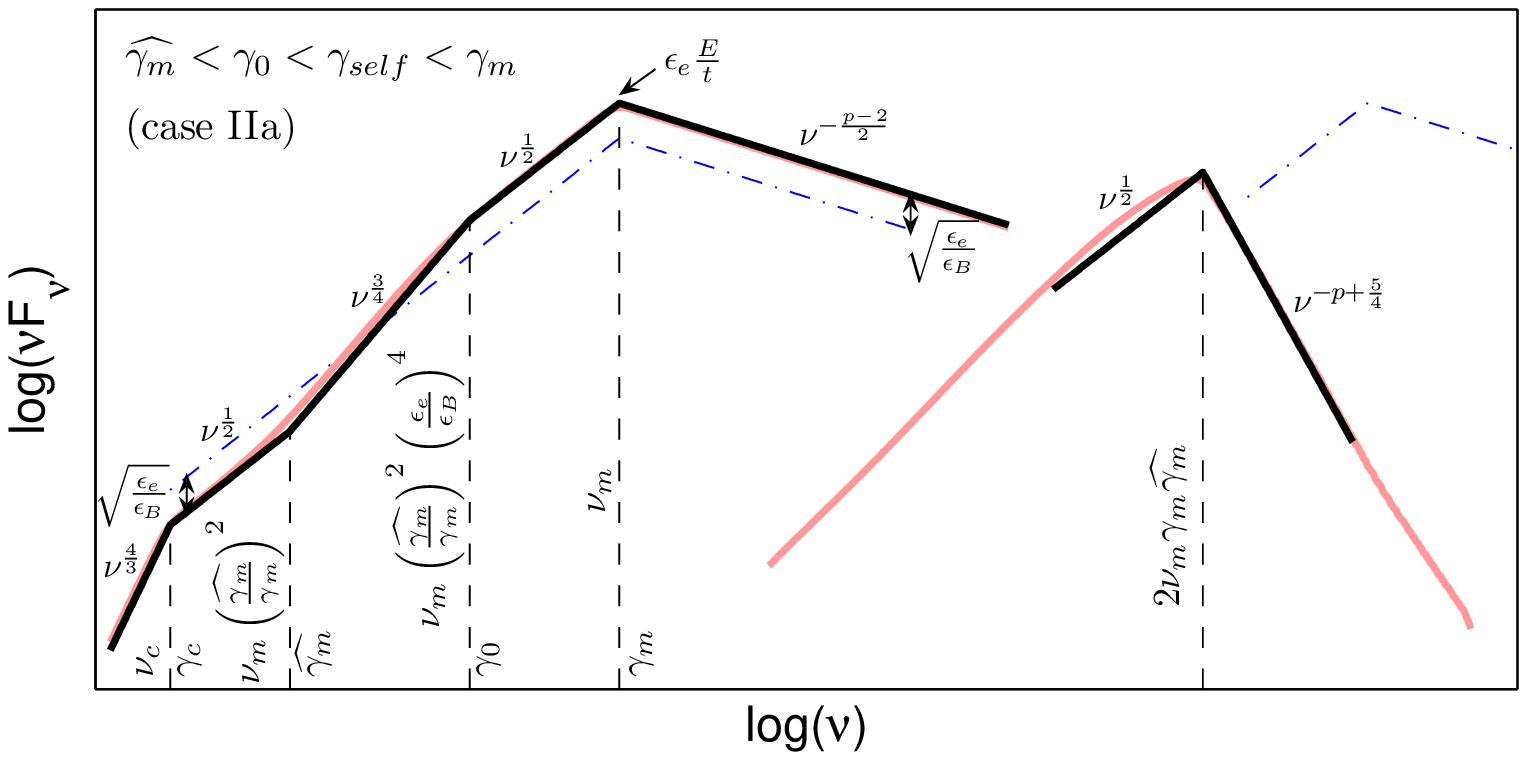}
  \caption{The synchrotron-SSC spectrum for $\g_c <\gmh<\g_0<\g_{self}<\g_m$
  (Case IIa). The specific parameters are
  $\g_m/\gmh=10^{11}$, $\epse/\epsB=10^{3}$ and $p=2.4$.
  The notations are similar to Fig. \ref{FIG case1}.}\label{FIG caseIIa}
\end{figure}

{\it SSC spectrum}: Most of the SSC energy is radiated by $\gamma_m$
electrons that upscatter synchrotron photons emitted by electrons
with $\gmt=\gamma_m^{1/2}\gmh^{1/2}$. Therefore, the peak of $\nu
F_{\nu}^{IC}$ is at $\nu_{peak}^{IC} \sim 2\nu_m \g_m \gmh$. The SSC
power-law index (of $F_\nu^{IC}$) at frequencies just below
$\nu_{peak}^{IC}$ can range between $-1/2$ and $-3/4$. At much lower
frequencies it can be as shallow as $-1/4$. At $\nu>\nu_{peak}^{IC}$
the power-law index range between $-p+1/2$ and $-p+1/4$ where at
higher frequencies ($\nu>2\nu_c \gch^2$) the index becomes $-p-1/3$.

An example of a case IIa analytic synchrotron-SSC spectrum and
comparison to the numerical spectrum is presented in figure \ref{FIG
caseIIa}. The normalization of the analytic SSC spectrum is
calculated using Eq. \ref{EQ Ygm_complex}.

%[Notes for my self: At very low frequencies most of the SSC energy
%is radiated by $\gamma_0$ electrons and then between $\nu_{KN,m}
%\gamma_0^2$ and $\nu_(\gamma_0) \gamma_0^2$ the spectral index is
%-1/4. later for $\nu > \nu_(\gamma_0) \gamma_{KN}^{-1}(\gamma_0)^2$
%the spectrum is $-3/4$ if this $\nu$ is smaller than
%$\nu_{peak}^{IC}$.]

\subsubsection{Case IIb: $\gmh<\g_{self}<\g_0<\g_m$
$~~\left[\left(\frac{\g_m}{\gmh}\right)^{1/3}<\reps<\frac{\g_m}{\gmh}\right]$}\label{SEC caseIIb}

In addition to $\nu_c$ and $\nu_m$ the break frequencies in this
case correspond to $\gmh$, $\goh$ and $\g_0$. $\widehat{\goh}$,
$\widehat{\gmh}$ and $\gch$ are all larger than $\g_0$ and therefore
do not affect the electron distribution (note that we assume
$\g_c<\gmh$). Here $\g_{self}$ electrons are cooling primely by
their own synchrotron photons and $\nu_{self}<\nu_{peak}^{syn}$. As
a result a new power-law segment is introduced. Plugging Eq. \ref{EQ
Y} into Eq. \ref{EQ Ng_fast} using $\g_{self}=\gst$, we get
$N_\g(\g_{self}) \propto \g^{\alpha(\g_{self})-1}$, which according
to Eq. \ref{EQ alpha} implies $\alpha(\g_{self})=0$. Therefore
$\alpha(\goh<\g<\g_0)=0$ and $Y(\goh<\g<\widehat{\goh})\propto
\g^{-1}$ implying:
\begin{equation}\label{EQ g0IIb}
   \g_0 \approx \sqrt{\reps \g_m \gmh}.
\end{equation}
Additional power-law segments are  $Y(\gmh<\g<\goh)\propto
\g^{-1/2}$ and $\alpha(\gmh<\g<\goh)=-1/4$ where the latter implies
$Y(\widehat{\gmh}<\g<\widehat{\goh})\propto \g^{-3/4}$. Similarly to
the previous case  $Y(\g<\gmh)=\epse/\epsB$ since $Y(\g_m)<1$. The
resulting $Y$ spectrum is therefore:
\begin{equation}\label{EQ Y2b}
    Y(\gamma) \approx \left\{\begin{array}{lc}
       \reps & \g< \gmh \\

       \reps \left(\frac{\g}{\gmh}\right)^{-1/2} &
       \gmh<\g<\g_m \left(\reps\right)^{-1}\\

       \left(\reps\right)^{1/2}\gmh^{1/2}\g_m^{1/2}\g^{-1} &
       \g_m \left(\reps\right)^{-1}<\g<\gmh \left(\reps\right)^{2}\\

       \g_m^{1/2}\gmh^{1/4}\g^{-3/4} & \gmh \left(\reps\right)^{2}<\g<\frac{\g_m^2}{\gmh} \\

       \gmh^{1/2}\g^{-1/2} & \frac{\g_m^2}{\gmh} <\g

       \end{array}\right. .
\end{equation}
The synchrotron spectrum in this case is affected only between
$\nuh$ and  $\nu_0$:
\begin{equation}\label{EQ Fsyn2b}
    F_\nu^{syn} \propto \left\{\begin{array}{lc}
        \nu^{-1/2} & \nu_c<\nu<\numh  \\
        \nu^{-1/4} & \numh <\nu<\nuoh\\
        \nu^{0} & \nuoh<\nu<\nu_0\\
        \nu^{-1/2} & \nu_0<\nu<\nu_m\\
        \nu^{-p/2} & \nu_m<\nu
                       \end{array}\right.
\end{equation}

\begin{figure}
\includegraphics[width=16.5cm]{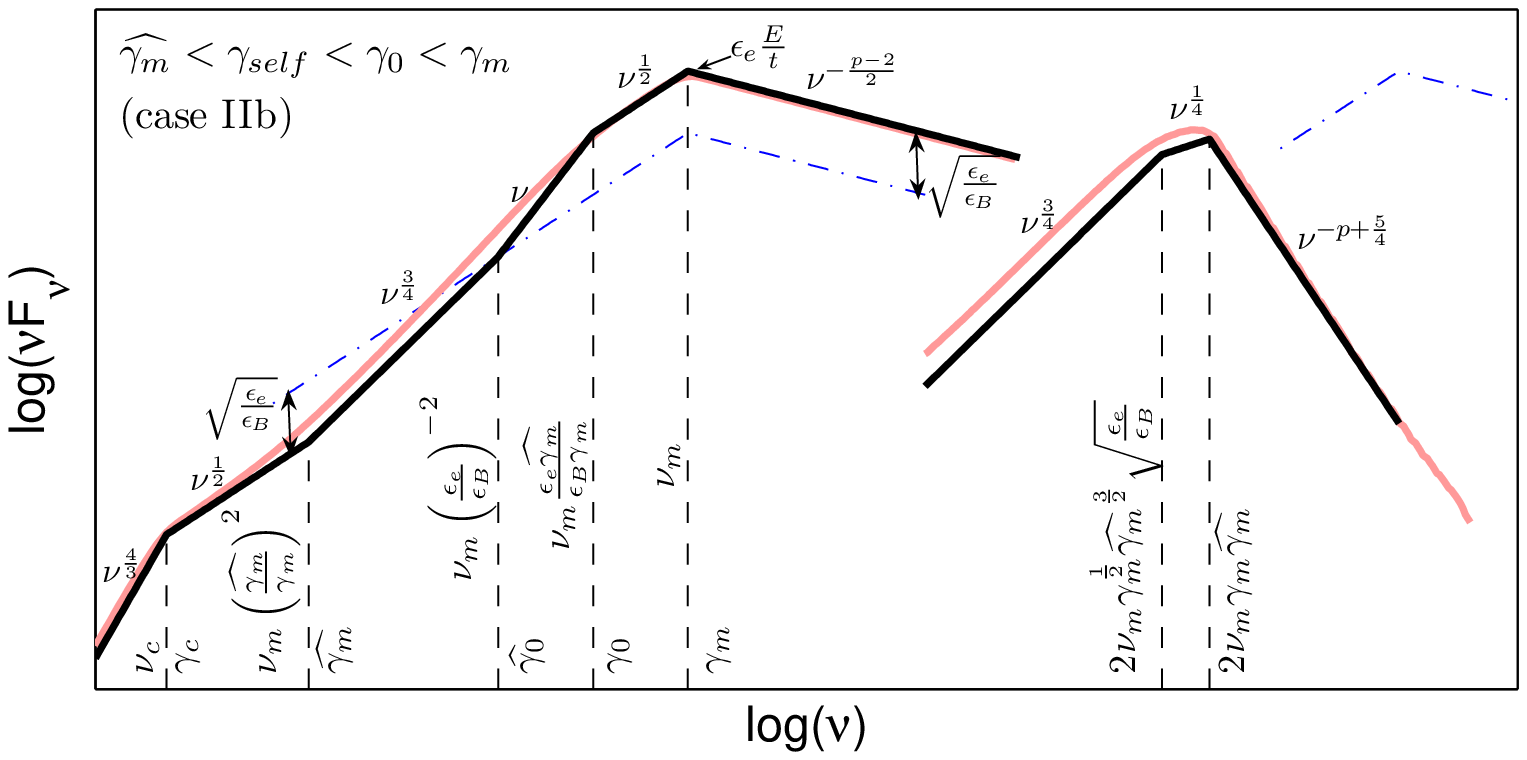}
  \caption{The synchrotron-SSC spectrum for $\g_c <\gmh<\g_{self}<\g_0<\g_m$
  (Case IIb). The specific parameters are
  $\g_m/\gmh=10^{8}$, $\epse/\epsB=10^{4}$ and $p=2.4$.
  The notations are similar to Fig. \ref{FIG case1}.}\label{FIG caseIIb}
\end{figure}

{\it SSC spectrum}: Similarly to the previous case  $\nu_{peak}^{IC}
\sim 2\nu_m \g_m \gmh$. Just below the peak the spectral index is
$-3/4$ down to $\sqrt{\frac{\gmh\epse}{\g_m\epsB}}\nu_{peak}^{IC}$
where it becomes $-1/4$.\footnote{for very large values of $\epse/\epsB
> 10^5$ it can approach $0$ before it return to $-1/4$.}
Above the peak the power-law index ranges between $-p$ to $-p+1/4$
where at higher frequencies it becomes $-p+1/2$ and at even higher
frequencies ($\nu>2\nu_c \gch^2$) it is $-p-1/3$.

An example of the analytic case IIb synchrotron-SSC spectrum and
comparison to the numerical spectrum is presented in figure \ref{FIG
caseIIb}. The normalization of the analytic SSC spectrum is
calculated using Eq. \ref{EQ Ygm_complex}.

\subsubsection{Case IIc: $\gmh<\g_{self}<\g_m<\g_0<\widehat{\gmh}$
$~~\left[\frac{\g_m}{\gmh}<\reps<\left(\frac{\g_m}{\gmh}\right)^3 \right]$}\label{SEC caseIIc}

Under these assumptions the critical Lorentz factors are $\gmh$,
$\goh$ and $\g_0$ (similarly to case IIb). In this case $\goh<\gmh$
and therefore $Y(\g) \propto \g^{-1}$ for $\gmh<\g<\widehat{\gmh}$.
Similarly to the previous case
\begin{equation}\label{EQ g0IIc}
   \g_0 \approx \sqrt{\reps \g_m \gmh}.
\end{equation}
Since $\g_m<\g_0$ the synchrotron power-law index above $\nu_m$ is
modified from $p/2$ to $(p-1)/2$ up to $\nu_0$, implying that if
$2<p<3$ then $\nu_{peak}^{syn}\approx\nu_0$ and as a result $Y(\g)
\propto \g^{(p-3)/2}$ for $\goh<\g<\gmh$. Note that $\g_0$ electrons
carry a fraction $\approx (\g_0/\g_m)^{2-p} $ of the total electrons
energy which they radiate entirely as synchrotron photons. Therefore
the maximal value of $Y$ when synchrotron emissivity is dominated by
$\g_0$ electrons (i.e., $p<3$) is $Y(\goh) \approx
\reps\left(\frac{\g_0}{\g_m}\right)^{2-p}$. Thus, the $Y$ spectrum
in case that $2<p<3$ is:
\begin{equation}\label{EQ Y2c}
    Y(\gamma) \approx \left\{\begin{array}{lc}
      \left(\reps\right)^{\frac{4-p}{2}}\left(\frac{\gmh}{\g_m}\right)^{\frac{2-p}{2}}&
      \g< \g_m \left(\reps\right)^{-1} \\

      \left(\reps\right)^\frac{1}{2}\g_m^{1/2}\gmh^{\frac{2-p}{2}}\g^{-\frac{3-p}{2}} &
      \g_m\left(\reps\right)^{-1}<\g<\gmh\\

      \left(\reps\right)^\frac{1}{2}\gmh^{1/2}\g_m^{1/2}\g^{-1} &
      \gmh<\g<\frac{\g_m^2}{\gmh}\\

      \left(\reps\right)^\frac{1}{2}\gmh^\frac{p+1}{4}\g_m^\frac{2-p}{2}\g^{-\frac{5-p}{4}}&
      \frac{\g_m^2}{\gmh}<\g<\left(\reps\right)^{2} \gmh \\

      \left(\reps\right)^\frac{p-2}{2}\gmh^\frac{p-1}{2}\g_m^\frac{2-p}{2}\g^{-1/2} &
      \left(\reps\right)^{2} {\gmh} <\g

                    \end{array}\right.
\end{equation}
The synchrotron spectrum in this case is affected only between
$\nuh_0$ and  $\nu_0$:
\begin{equation}\label{EQ Fsyn2c}
    F_\nu^{syn} \propto \left\{\begin{array}{lc}
       \nu^{-\frac{1}{2}} & \nu_c<\nu<\nuoh \\
       \nu^{-\frac{p-1}{4}} &  \nuoh<\nu<\numh\\
       \nu^{0} & \numh<\nu<\nu_m\\
       \nu^{-\frac{p-1}{2}} & \nu_m<\nu<\nu_0\\
       \nu^{-\frac{p}{2}} & \nu_0<\nu
                       \end{array}\right.
\end{equation}
In case that $p>3$ most of the synchrotron energy is emitted at
$\nu_m$, eliminating the second and fourth power-law segments in Eq.
\ref{EQ Y2c}. Instead, $Y(\g<\gmh)= [(\epse\g_m)/(\epsB\gmh)]^{1/2}$
and $Y(\g>\g_m^2/\gmh) \propto \g^{-1/2}$. Eq. \ref{EQ Fsyn2c} is
then revised so $F_\nu \propto \nu^{1/2}$ for all $\nu_c<\nu<\nu_m
(\gmh/\g_m)^2$.

If $\widehat{\gmh}<\g_0$ then $\widehat{\gmh}$ becomes a new
critical lorentz factor and two more power-law segments are added.
slightly modifying Eqs. \ref{EQ Y2c}--\ref{EQ Fsyn2c}. Yet more
power-law segments are added if $\widehat{\widehat{\gmh}}<\g_0$ and
so on. Asymptotically the spectrum approaches the case where
$\g_m=\gmh$ which we solve next (case III). Moreover, the criterion
for  $\widehat{\gmh}<\g_0$ is $\g_m/\gmh<(\epse/\epsB)^{1/3}$ which
for most typical values of $(\epse/\epsB)$ implies $\g_m/\gmh
\lesssim 10$ and therefore this case (as well as all higher order
cases) is well approximated by case III.

\begin{figure}
\includegraphics[width=16.5cm]{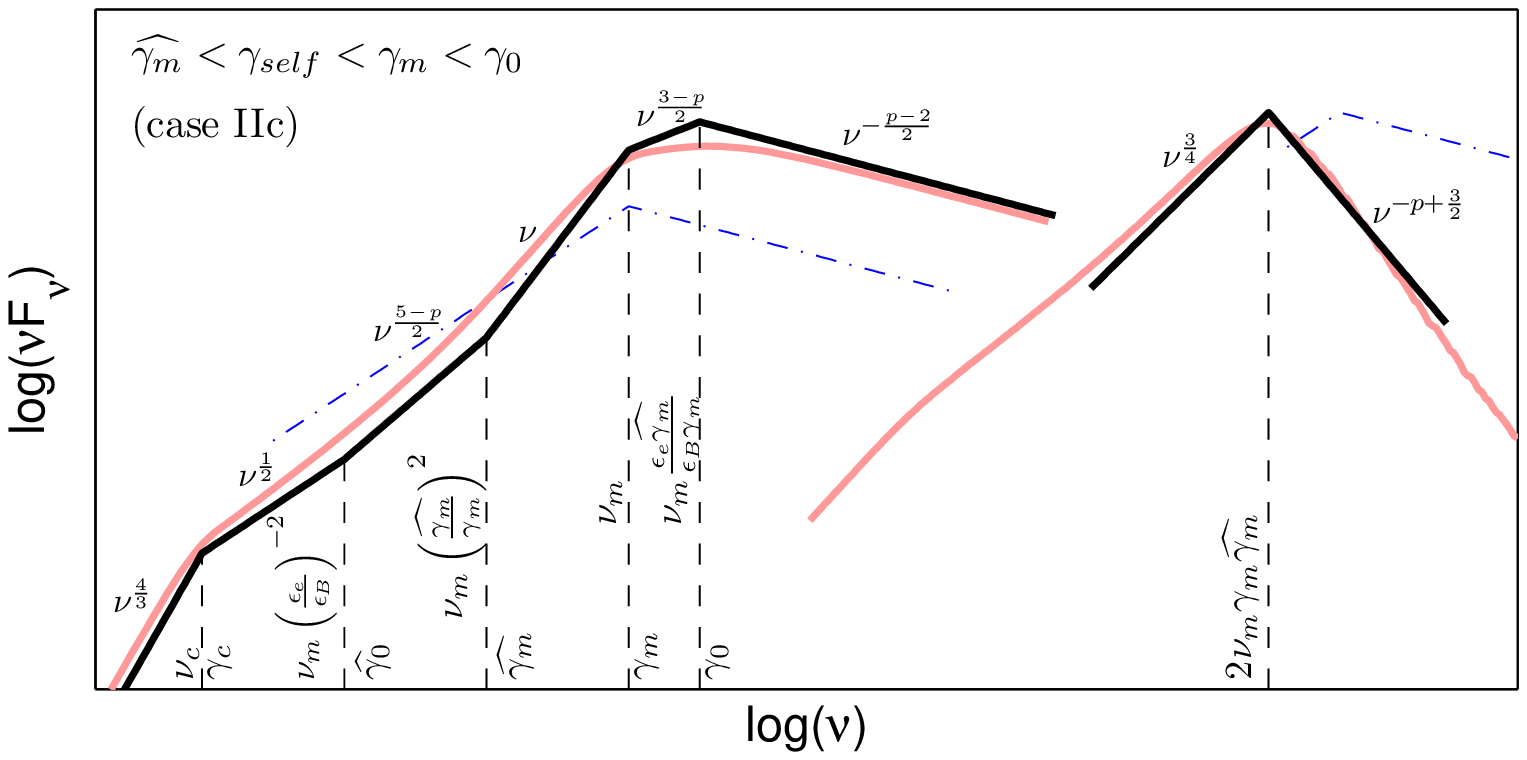}
  \caption{The synchrotron-SSC spectrum for $\g_c <\gmh<\g_{self}<\g_m<\g_0$
  (Case IIc). The specific parameters are
  $\g_m/\gmh=10^{2}$, $\epse/\epsB=10^{4}$ and $p=2.4$.
  The notations are similar to Fig. \ref{FIG case1}.}\label{FIG caseIIc}
\end{figure}

{\it SSC spectrum}: Similarly to the two previous cases
$\nu_{peak}^{IC} \sim \nu_m \g_m \gmh$. The spectral index below the
peak ranges between  $-1/4$ and $0$ (where the latter can be
observed only for very large values of $\epse/\epsB
> 10^5$) and above the peak it ranges between $-p$ and $-p+1/2$
where at higher frequencies ($\nu>2\nu_c \gch^2$) it becomes
$-p-1/3$.

An example of the analytic case IIc synchrotron-SSC spectrum and
comparison to the numerical spectrum is presented in figure \ref{FIG
caseIIc}. The normalization of the analytic SSC spectrum is
calculated using Eq. \ref{EQ Ygm_complex}.

\subsection{Case III - $\g_m=\gmh$}\label{SEC caseIII}

Here $\g_m=\gmh=\widehat{\gmh}=...$ and therefore $\g_0$ and
potentially $\goh$ are the only critical Lorentz factors in addition
to $\g_c$ and $\g_m$. As we show below there is a slight difference
between cases where $p<2.5$, for which $\nu_{peak}^{syn}=\nu_0$, and
spectra with $p>2.5$ where $\nu_{peak}^{syn}=\nu_m$. In both cases
electrons with $\g_m<\g<\g_0$ are cooling predominantly by
upscattering photons with $\nu<\nu_m$ but if
$\nu_{peak}^{syn}=\nu_0$ then electrons with $\g<\g_m$ are cooling
by upscattering $\nu>\nu_m$ photons while if
$\nu_{peak}^{syn}=\nu_m$ then they are cooling by upscattering
$\nu_m$ photons. Solving for the mutual dependence of the spectral
slopes above and below $\nu_m$ on each other in case that
$\nu_{peak}^{syn}=\nu_0$ (using Eqs. \ref{EQ Ng_fast}, \ref{EQ Y}
and \ref{EQ Fnusyn_general}) results in
$\alpha(\goh<\g<\g_m)=(1-p)/3$ and $\alpha(\g_m<\g<\g_0)=2(1-p)/3$
\citep[see also ][]{Peer05b}. This result show that the transition
from $\nu_{peak}^{syn}=\nu_0$ to $\nu_{peak}^{syn}=\nu_m$ takes
place at p=2.5.
\begin{figure}
\includegraphics[width=16.5cm]{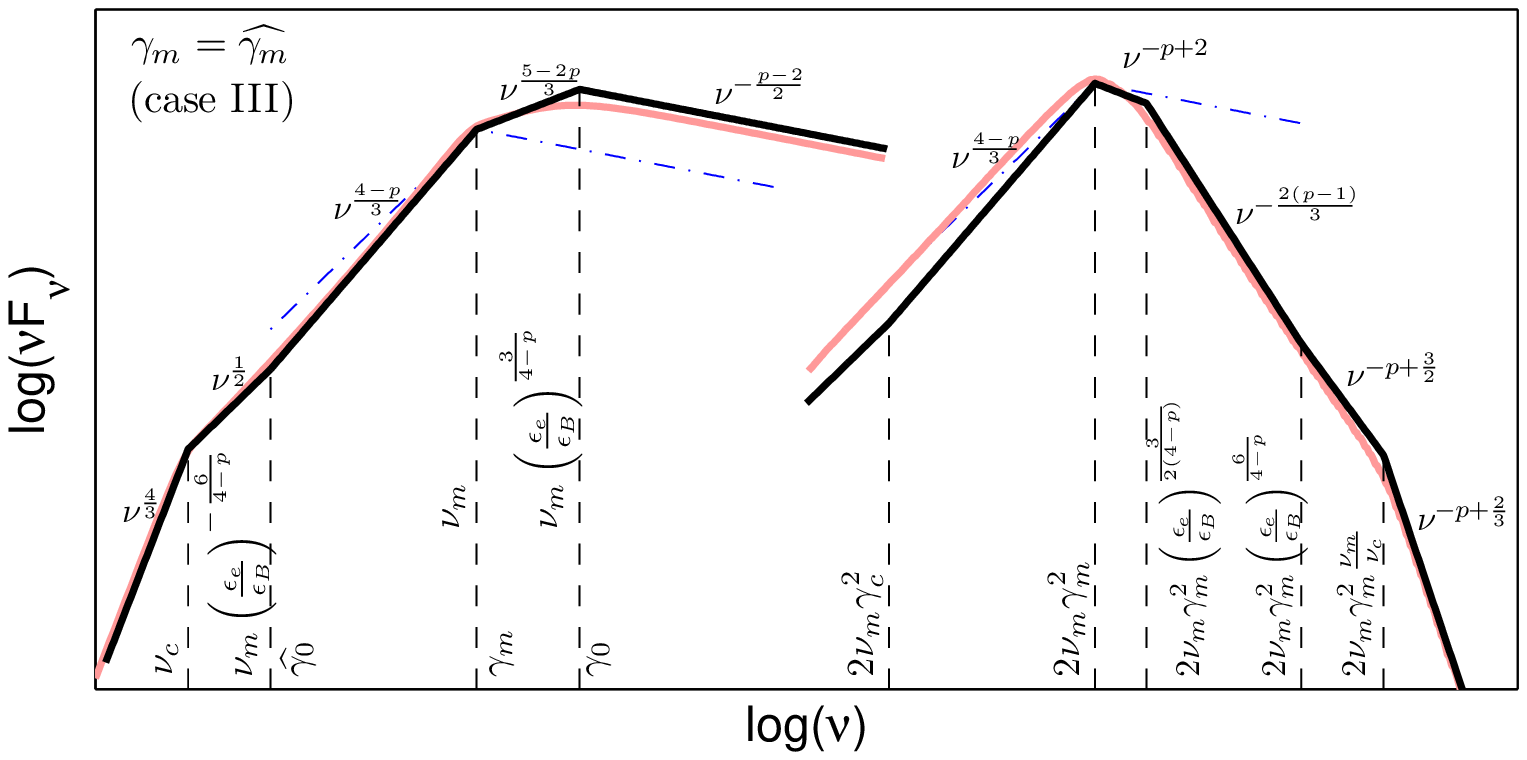}
  \caption{The synchrotron-SSC spectrum for $\g_c<\gmh=\g_m$
  (Case III). The specific parameters are
  $\epse/\epsB=10^{3}$ and $p=2.2$.
  The notations are similar to Fig. \ref{FIG case1}.}\label{FIG caseIII}
\end{figure}

Therefore if $p<2.5$ then:
\begin{equation}\label{EQ g0III}
\g_0=\g_m(\epse/\epsB)^\frac{3}{2(4-p)} ~~;~~ p<2.5,
\end{equation}
The $Y$ spectrum is
\begin{equation}\label{EQ Y3}
    Y(\gamma) \approx \left\{\begin{array}{lc}
      \left(\reps\right)^{\frac{1}{2}+\frac{5-2p}{4-p}}&
      \g< \g_m \left(\reps\right)^{-\frac{3}{4-p}} \\

      \left(\reps\right)^\frac{1}{2}\left(\frac{\g }{\g_m}\right)^{-\frac{5-2p}{3}} &
      \g_m \left(\reps\right)^{-\frac{3}{4-p}} <\g<\g_m\\

      \left(\reps\right)^\frac{1}{2}\left(\frac{\g }{\g_m}\right)^{-\frac{4-p}{3}} &
      \g_m<\g<\g_m \left(\reps\right)^{-\frac{6}{4-p}}\\

      \left(\reps\right)^\frac{3(p-2)}{2(4-p)}\left(\frac{\g }{\g_m}\right)^{-\frac{1}{2}} &
      \g_m \left(\reps\right)^{\frac{6}{4-p}}<\g<\gch\\

                    \end{array}\right. ~~;~~ p<2.5,
\end{equation}
the synchrotron spectrum is:
\begin{equation}\label{EQ Fsyn3}
    F_\nu^{syn} \propto \left\{\begin{array}{lc}
       \nu^{-\frac{1}{2}} & \nu_c<\nu<\nuoh \\
       \nu^{-\frac{p-1}{3}} &  \nuoh<\nu<\nu_m\\
       \nu^{-\frac{2(p-1)}{3}} & \nu_m<\nu<\nu_0\\
       \nu^{-\frac{p}{2}} & \nu_0<\nu
                       \end{array}\right. ~~;~~ p<2.5,
\end{equation}
and the SSC spectrum is:
\begin{equation}\label{EQ FIC3}
    F_\nu^{IC} \propto \left\{\begin{array}{lc}
       \nu^{-\frac{1}{2}} & 2\nu_m\g_c^2<\nu<2\nu_{m}\goh^2\\
       \nu^{-\frac{p-1}{3}} &  2\nu_m\goh^2<\nu<2\nu_m\g_m^2\\
       \nu^{-p+1} &  2\nu_m\g_m^2<\nu<2\nu_m\g_m \g_0\\
       \nu^{-\frac{2p+1}{3}} & 2\nu_m\g_m \g_0<\nu<2\nu_m\g_m \widehat{\goh}\\
       \nu^{-p+\frac{1}{2}} & 2\nu_m\g_m \widehat{\goh}<\nu
                       \end{array}\right. ~~;~~ p<2.5 .
\end{equation}

In case that $p>2.5$ then Eq. \ref{EQ Y3} can be used by
substituting $p\rightarrow 2.5$. The synchrotron spectrum (Eq
\ref{EQ Fsyn3}) is modified so the spectral index at
$\nu_c<\nu<\nu_m$ is $-1/2$ and at $\nu_m<\nu<\nu_0$ it is
$-p/2+1/4$, where $\g_0=\g_m(\epse/\epsB)$. The spectral index of
SSC spectrum at $2\nu_c \g_c^2<\nu<2\nu_m\g_m^2$ is $-1/2$, at
$2\nu_m\g_m^2<\nu<2\nu_m\g_m\g_0$ it is $-p+1$ and at
$\nu>2\nu_m\g_m\g_0$ it is $-p+1/2$.

An example of the analytic case III synchrotron-SSC spectrum with
$p<2.5$ and comparison to the numerical spectrum is presented in
figure \ref{FIG caseIII}. The normalization of the analytic SSC
spectrum is calculated using Eq. \ref{EQ Ygm_complex}.

\subsection{The dependence of the SSC to synchrotron luminosity
ratio on $\g_m/\gmh$}\label{SEC YbarFast}

The SSC to synchrotron luminosity ratio is an observable that can be
used in order to constrain the physical parameters of the source.
Here we derive this ratio as a function of $\g_m/\gmh$, where the
rest of the physical parameters (e.g., $\epse/\epsB$) are held
constant and $\g_c \ll \g_m,\gmh$, so the value of $\g_c$ do not
affect the results. The SSC to synchrotron luminosity ratio is an
average of $Y(\gamma)$ weighted by the synchrotron emissivity:
\begin{equation}\label{EQ Ybar}
    \overline{Y}\equiv \frac{L_{IC}}{L_{syn}}=\frac{\int Y P_{syn}
    N_\g d\g} {\int P_{syn}
    N_\g d\g} .
\end{equation}
In case where KN effects can be neglected then $Y$ is a constant and
$\overline{Y}=Y_{noKN}$ (see Eqs. \ref{EQ YnoKN}, \ref{EQ YnoKN2}).
The SSC luminosity is dominated by $\g_m$ electrons. Typically, the
synchrotron luminosity is dominated by $\g_m$ electrons as well in
which case $\overline{Y} \approx Y(\g_m)$. The value $Y(\g_m)$ is
also of interest since there are cases where $F_{\nu,syn}(\g_m)$ is
the observeable. Therefore, we first find $Y(\g_m)$.

The dependence of $Y(\g_m)$  on the ratio $\g_m/\gmh$ can be
approximated using Eqs. \ref{EQ Y2a}, \ref{EQ Y2b} and \ref{EQ Y2c}.
For $\g_m/\gmh \ll 1 $ there is no significant KN effect and
$Y(\g_m) \approx (\epse/\epsB)^{1/2}$. When $\g_m/\gmh \approx 1$
the energy in $\g_m$ electrons is roughly a fraction of $\approx
p-2$ out of the total energy in electrons and therefore $Y(\g_m)
\approx [(p-2)\epsilon_e/\epsilon_B]^{1/2}$. For $\g_m/\gmh \lesssim
1$ (case IIc) and $\g_m/\gmh \ll 1$ (part of case IIa) one obtains
$Y(\g_m) \propto (\g_m/\gmh)^{-1/2}$, with a slightly different
dependence ($\propto (\g_m/\gmh)^{-1/4}$) for intermediate values of
$\g_m/\gmh$. Therefore, a simple approximation for $Y(\g_m)$ is
\citep{Ando08}:
\begin{equation}\label{EQ Ygm}
    Y(\gamma_m) \approx \sqrt{\reps}\left\{\begin{array}{lc}
    1 & \frac{\g_m}{\gmh}<p-2\\
    \sqrt{p-2}\left(\frac{\g_m}{\gmh}\right)^{-1/2} &\frac{\g_m}{\gmh}>p-2
                                \end{array}\right.
\end{equation}
\begin{figure}
  \includegraphics[width=16cm]{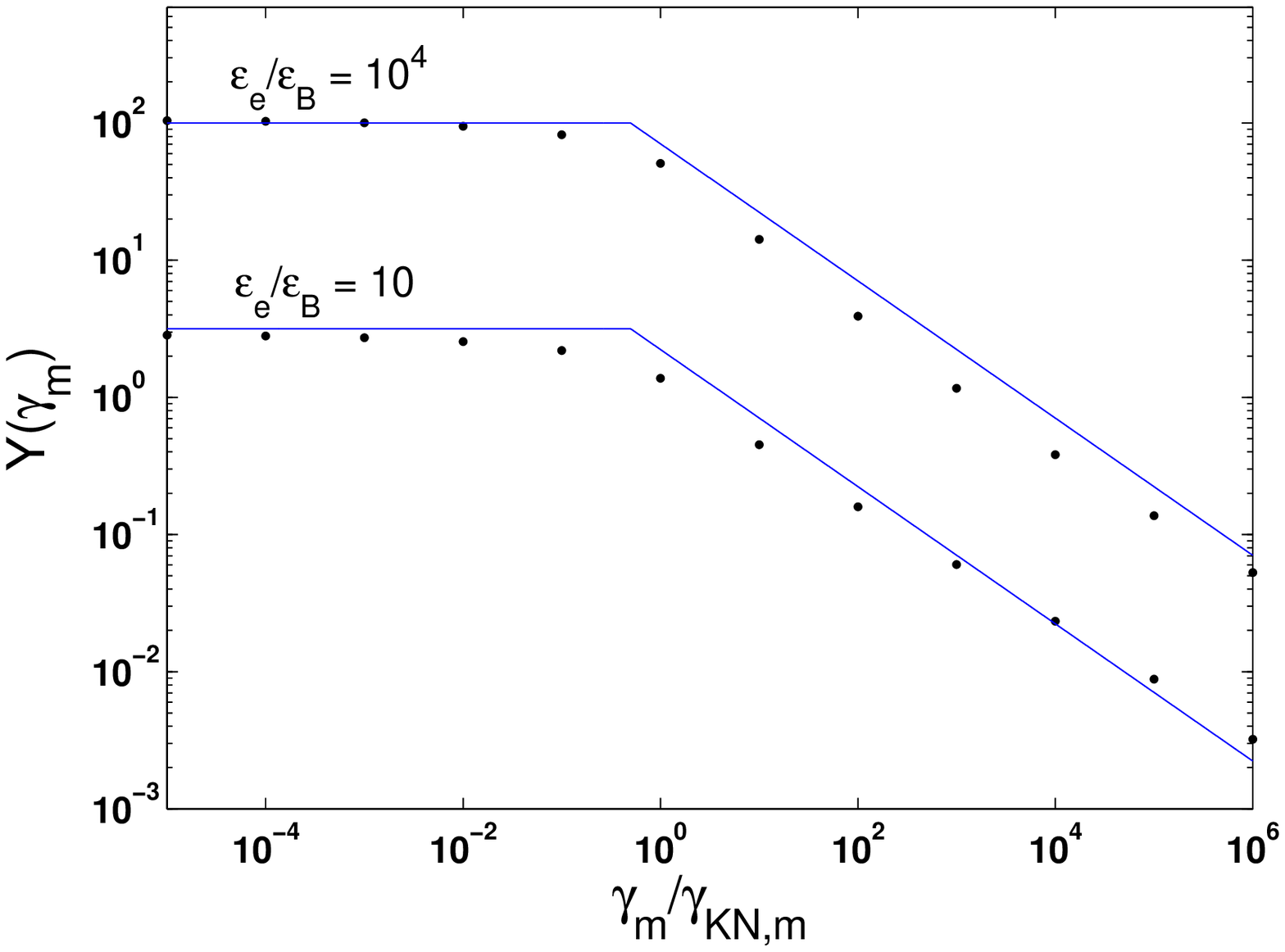}
  \caption{The value of $Y(\g_m)$ as calculated numerically by integrating Eqs.
  \ref{EQ cool func}, \ref{EQ dN_dt} and \ref{EQ Ynumeric} ({\it dots})
  compared to the simple approximation of Eq. \ref{EQ Ygm}
  ({\it solid lines}) for two values of $\epse/\epsB$ and $p=2.5$ .}\label{FIG Ym1}
\end{figure}
The quality of this approximation is depicted in Fig. \ref{FIG Ym1}.
It is accurate to within a factor of $\approx 2$ as long as
$\g_m/\gmh < (\epse/\epsB)^2$ or for any reasonable value of
$\g_m/\gmh$ when $\epse/\epsB \lesssim 100$. A more accurate
approximation which include the intermediate segment of $Y \propto
\g^{-1/4}$ and should be used for large values of $\epse/\epsB$ is:
\begin{equation}\label{EQ Ygm_complex}
    Y(\g_m) \approx \left\{\begin{array}{lc}
    \sqrt{\reps} & \frac{\g_m}{\gmh}<p-2\\
    \sqrt{\frac{(p-2)\epse}{\epsB}}\left(\frac{\g_m}{\gmh}\right)^{-1/2}
    & p-2<\frac{\g_m}{\gmh}<\left(\reps\right)^2\\
    \sqrt{p-2}\left(\frac{\g_m}{\gmh}\right)^{-1/4}
    & \left(\reps\right)^2<\frac{\g_m}{\gmh}<\left(\reps\right)^4\\
    \sqrt{p-2}\reps\left(\frac{\g_m}{\gmh}\right)^{-1/2}
    & \left(\reps\right)^4<\frac{\g_m}{\gmh}
    \end{array}\right.
\end{equation}

\begin{figure}
\includegraphics[width=16cm]{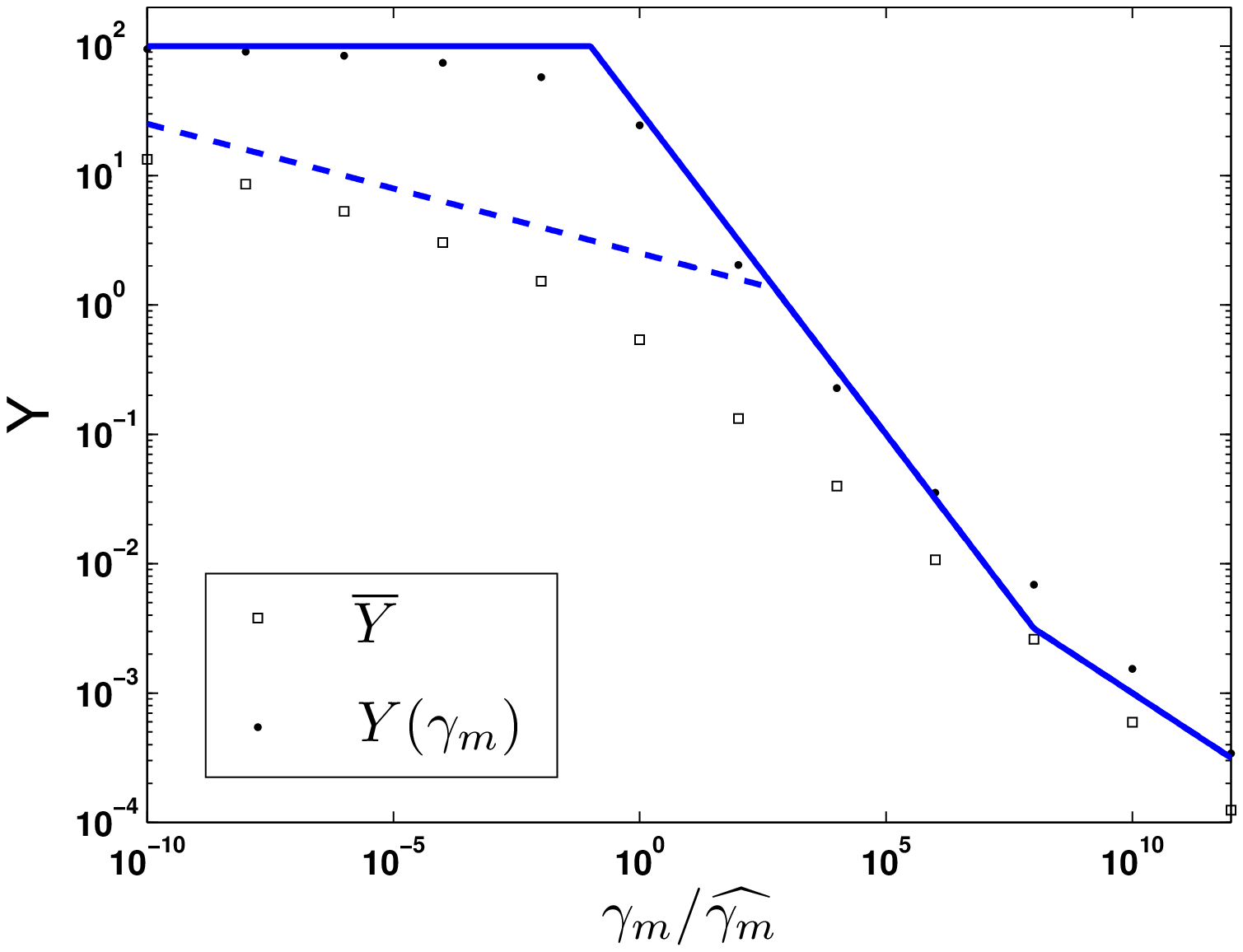}
  \caption{The value of $\overline{Y}$ ({\it open squares}) and
  $Y(\gamma_m)$ ({\it dots}) as calculated numerically (by integrating Eqs.
  \ref{EQ cool func}, \ref{EQ dN_dt} and \ref{EQ Ynumeric}), compared to
  the analytic approximation Eq. \ref{EQ Ybar_approx} ({\it dashed lines})
  and Eq. \ref{EQ Ygm_complex} ({\it solid lines}).
  The parameters are $p=2.1$ and $\epsilon_e/\epsilon_B = 10^4$.}\label{FIG Ybar}
\end{figure}

The approximation $\overline{Y} \approx Y(\g_m)$ is good as long as
the synchrotron luminosity peaks at $\nu_m$. This is not the case
when $p<2.5$ and
$(\epse/\epsB)^{((p-2.5)/(p-2))}<\g_m/\gmh<\epse/\epsB$, where the
synchrotron luminosity is dominated by $\g_0$ electrons and
$\overline{Y} \approx Y(\gamma_m)
L_{syn}(\gamma_m)/L_{syn}(\gamma_0)$ (note that the SSC luminosity
is dominated by $\g_m$ electrons also in this regime). Since $\g_0
\approx \g_m (\epse/\epsB)$ for $\gamma_m<\gmh$, a reasonably simple
approximation for $\overline{Y}$ is:
\begin{equation}\label{EQ Ybar_approx}
    \overline{Y} \approx \left\{\begin{array}{lc}
    \left(\frac{\epsB\g_m}{\epse\gmh}\right)^{p-2}
    & ~~{\rm if}~~ \left(\reps\right)^{-\frac{2.5-p}{p-2}}<
    \frac{\g_m}{\gmh}<\reps(p-2)^\frac{1}{2(2.5-p)}~ \&~ p<2.5\\
    Y(\gamma_m) & {\rm otherwise}
    \end{array}\right. .
\end{equation}
Fig. \ref{FIG Ybar} shows this approximation for large value of
$\epse/\epsB=10^4$ and $p=2.1$  where the approximation
$\overline{Y}=Y(\g_m)$ is not adequate. Eq. \ref{EQ Ybar_approx}
provides a reasonable approximation that is accurate only to within
an order of magnitude. The largest deviation of Eq. \ref{EQ
Ybar_approx} is in the range $\gmh<\g_m<\gmh(\epse/\epsB)$ where it
can overestimate the  value of $\overline{Y}$ by an order of
magnitude.

\subsection{Dominant synchrotron cooling [$\epse\lesssim \epsB$]}\label{SEC dominantsynch fast}
Since only the electrons' energy can be radiated, $U_{ph}<\epsilon_e
U$. Therefore, $\epsilon_e<\epsilon_B$ guarantees that the magnetic
field energy density is larger than the synchrotron photon energy
density, and the cooling of all the electrons is dominated by
synchrotron emission. In such a case, the synchrotron spectrum is
not affected at all by the SSC emission. The SSC spectrum, however,
is affected by the KN limit:
\begin{equation}\label{EQ FIC synchcool}
    F_\nu^{IC} \propto \left\{\begin{array}{lc}
       \nu^{-\frac{1}{2}} & 2\nu_c\g_c^2<\nu<2\nu_{m}\g_m \min\{\g_m,\gmh\}\\
       \nu^{-p+\frac{1}{2}} &  2\nu_m\g_m \min\{\g_m,\gmh\}<\nu<2\nu_c\gch^2\\
       \nu^{-(p+\frac{1}{3})} &  2\nu_c\gch^2<\nu\\
       \end{array}\right. .
\end{equation}
The SSC to synchrotron energy output is well approximated by
$Y(\g_m)$ and it is:
\begin{equation}\label{EQ Y synchcool}
    \overline{Y} \approx Y(\gamma_m) \approx \reps\left\{\begin{array}{lc}
    1 & \frac{\g_m}{\gmh}<(p-2)^2\\
    (p-2)\left(\frac{\g_m}{\gmh}\right)^{-1/2}
    &\frac{\g_m}{\gmh}>(p-2)^2
    \end{array}\right. ,
\end{equation}
where we assume here $\g_c < \gmh$.

\section{Slow cooling}\label{SEC slowcooling}
This regime is more simple since the electron distribution is not
affected by inverse Compton (or synchrotron) cooling at $\g < \g_c$
while SSC cooling of electrons with $\g > \g_c$ is always dominated
by upscattering synchrotron photons with frequency $\leq \nu_c$.
This significantly simplifies the electron distribution. We present
in this regime only the case that $\g_c<\gmh$ since if $\g_c>\gmh$
then $Y(\g_c) <1$ (see \S\ref{SEC g_c}) and SSC cooling has no
effect on the electron distribution. We also discuss only the case
of $2<p<3$ which implies that $\nu_{peak}^{syn} \gtrsim \nu_c$ (for
$p>3$ the peak frequency is,in most cases, $\nu_m$).

Under these assumptions $Y$ typically takes the form (see exception
below):
\begin{equation}\label{EQ. Yslow}
    Y(\gamma)=Y(\gch) \left\{\begin{array}{lc}
      1 & \g < \gch\\
      \left(\frac{\g}{\gch}\right)^\frac{p-3}{2} & \gch<\g < \gmh \\
      \left(\frac{\g_c}{\g_m}\right)^{p-3}\left(\frac{\g}{\gmh}\right)^{-\frac{4}{3}} &  \gmh<\g
      \end{array}\right.
\end{equation}
The value of $Y(\gch)$ can be found by the normalization at $\g_c$:
\begin{equation}\label{EQ. Yc slow}
    Y(\g_c)[1+Y(\g_c)] \approx \reps\left(\frac{\g_c}{\g_m}\right)^{2-p}\left(\frac{ \min\{\g_c,\gch \}}{\g_c}\right)^{\frac{3-p}{2}}.
\end{equation}
In the case of $\g_c \ll \gch$ and $Y(\g_c)>1$, Eq. \ref{EQ. Yc
slow} is reduced to the slow cooling value in case that KN effects
are neglected, $Y=(\epse/\epsB)^{1/2}(\g_c/\g_{m})^{(2-p)/2}$
\citep[e.g.,][]{Sari01}. Equation \ref{EQ. Yc slow} can be used also
to find $\g_c$  when KN effects play an important role (see
\S\ref{SEC g_c}).

The synchrotron spectrum above $\nu_c$ is:
\begin{equation}\label{EQ FsynSlow1}
    F_{\nu}^{syn} \propto \left\{\begin{array}{lc}

      \nu^{-p/2} &  \nu_c<\nu<\nuch \\

      \nu^{-\frac{3}{4}(p-1)} &
      \max\{\nuch,\nu_c\}<\nu<\min\{\numh,\nu_0\}\\

      \nu^{-(p/2-2/3)} & \numh<\nu<\nu_0
      ~;~{\rm only ~if~ \gmh<\g_0} \\

      \nu^{-p/2} &  \nu_0<\nu
    \end{array}\right.,
\end{equation}
Since the spectrum is altered only at $\nu>\max\{\nu_c,\nu_0\}$ not
all these segments exist in all cases. For example, the first
power-law segment can be observed only if $\g_c<\gch$ while the
third segments can be observed only if $\gmh<\g_0$. The value of
$\g_0$ can be calculated using Eqs. \ref{EQ. Yslow} and \ref{EQ. Yc
slow}. We give here the value of $\g_0$ for the case $\g_c<\gch$
which is the most likely to have an observable signature:
\begin{equation}\label{EQ g0 slow}
    \g_0=\left\{\begin{array}{lr}
      \gch\left(\frac{\g_c}{\g_m}\right)^\frac{p-2}{3-p}\left(\reps\right)^\frac{1}{3-p} &  ~~~~\reps<\left(\frac{\g_c}{\g_m}\right)^{4-p} \\
      \gmh\left(\frac{\g_c}{\g_m}\right)^\frac{3(p-4)}{8}\left(\reps\right)^\frac{3}{8} &  ~~~~\reps>\left(\frac{\g_c}{\g_m}\right)^{4-p}
                \end{array}\right. ,
\end{equation}
where $\reps<\left(\frac{\g_c}{\g_m}\right)^{4-p}$ is the condition
for $\g_0<\gmh$.

When $p - 2 \ll 1$ and $\nu_{peak}^{syn}>\nu_c$, then there is
another power-law segment for $Y(\g<\g_c \cdot \min\{1,\g_c/\gch\})$
where $Y \propto \gamma^{(3p-7)/4}$. Yet another segment $Y \propto
\g^{(3p-10)/2}$, corresponding to the third power-law segment in Eq.
\ref{EQ FsynSlow1}, exist in case that the synchrotron power peaks
at $\nu>\numh$. These additional segments affect only $\g < \g_c$
and therefore there is no farther effect on the electrons
distribution or the synchrotron spectrum.

The KN limit affect the SSC spectrum only at $\nu>2\nu_c\gch\cdot
\max\{\gch,\g_c\}$. This regime is therefore rather similar to case
I of the fast cooling regime and the same line of reasoning we used
there results in:
\begin{equation}\label{EQ FIC slow}
    F_{\nu}^{IC} \propto \left\{\begin{array}{lc}

      \nu^{-p/2} & 2\nu_c\g_c^2<\nu<2\nu_c\gch^2\\

      \nu^{-(p-1)} &  2\nu_c\gch\cdot \max\{\gch,\g_c\} <\nu<2\nu_c\gch\cdot \min\{\g_0,\gmh\}\\

      \nu^{-\frac{p+1}{2}} & 2\nu_c\gch\g_0<\nu<2\nu_c\gch\gmh \\

      \nu^{-(p+\frac{1}{3})} &  2\nu_c \gch \cdot \max\{\g_0,\gmh\}<\nu
    \end{array}\right. .
\end{equation}
Not all these segments are always observed. The first segment is
observed only when $\g_c<\gch$ and the third is observed only if
$\g_0<\gmh$. In case that $\g_c<\gch$ the total SSC luminosity is
not significantly suppressed by the KN limit and the SSC peak is
observed at $\nu_{peak}^{IC} \approx 2\nu_c \g_c^2$. If however,
$\g_c>\gch$ the SSC peak is observed at $\nu_{peak}^{IC} \approx
2\nu_c \g_c \gch$ and the SSC luminosity (i.e., $\overline{Y}$) is
suppressed. The value of $\overline{Y}$ can be approximated by
$Y(\g_c)$ (Eq. \ref{EQ. Yc slow}) following the same reasoning
explained in the fast cooling case (\S \ref{SEC YbarFast}). Here the
approximation is very good for $p > 2.5$ where for lower values of
$p$ the approximation is an overestimate ($Y(\g_c)>\overline{Y}$)
when the synchrotron luminosity is dominated by $\g_0$ electrons. In
this case a better approximation is $\overline{Y} \approx Y(\g_c)
L_{syn}(\g_c)/L_{syn}(\g_0)$, which can be calculated for a given
set of parameters using Eqs. \ref{EQ. Yslow}-\ref{EQ FsynSlow1}.

\subsection{Dominant synchrotron cooling [$\epse\lesssim \epsB(\g_m/\g_c)^{p-2}$]}\label{SEC dominantsynch slow}
When synchrotron cooling is dominant also in the Thomson regime then
the synchrotron spectrum is not affected by IC scattering while the
SSC spectrum is:
\begin{equation}\label{EQ FIC Slowsynchcool}
    F_\nu^{IC} \propto \left\{\begin{array}{lc}
       \nu^{-\frac{1}{2}} & 2\nu_m\g_m^2<\nu<2\nu_{c}\g_c \min\{\g_c,\gch\}\\
       \nu^{-\frac{p+1}{2}} &  2\nu_c\g_c \min\{\g_c,\gch\}<\nu<2\nu_m\gmh^2\\
       \nu^{-(p+\frac{1}{3})} &  2\nu_m\gmh^2<\nu\\
       \end{array}\right. .
\end{equation}
$Y(\g_c)$ is a good approximation of $\overline{Y}$ in this regime
and it follows Eq. \ref{EQ. Yc slow}.

\section{The cooling frequency}\label{SEC g_c}

In sections \S\ref{SEC fastcooling} and \S\ref{SEC slowcooling} we
have described the synchrotron and SSC spectra given
$\epsilon_e/\epsilon_B$, $\gamma_m$, $\hat \gamma$ and the cooling
Lorentz factor $\gamma_c$. However, the value of $\g_c$ is also
affected by the KN limit and should be solved
self-consistently.\footnote{ The value of $\g_m$ is not affected by
the KN limit and can be found, e.g., in \cite{Sari98}}. In this
section, we show how to solve for $\gamma_c$. Moreover, we assumed
before that $\nu_c$ is sufficiently low. We discuss here the
modification to the spectra if $\nu_c$ is not low enough.

Electrons are cooling fast, i.e. radiating a large fraction of their initial
energy over the lifetime of the system, if their Lorentz factor
satisfies:
\begin{equation}\label{EQ cooling criterion}
    \g[1+Y(\g)] >\gamma_c^{syn} \equiv  \frac{6 \pi m_e c}{\sigma_T B^2 t} ,
\end{equation}
here, $\gamma_c^{syn}$ is the cooling Lorentz factor if SSC cooling
is neglected altogether  \citep[e.g.,][]{Sari98}. When SSC cooling
is taken into account, but KN effects are neglected $Y=Y_{noKN}$ is
independent of $\g$. Therefore, the l.h.s of the equation always
increase monotonically with $\g$ so that
$\g_c=\g_c^{syn}/(1+Y_{noKN})$ is defined as a critical cooling
frequency such that all electrons with $\g>\g_c$ are cooling fast
and all electrons with $\g<\g_c$ are not cooling over the system
lifetime. When KN effects are taken into account, $Y$ depends on
$\g$ and since there are cases where $\g Y(\g)$ is decreasing, e.g.
$Y(\g) \propto \g^{-4/3}$, the equation
\begin{equation}\label{EQ g_c general}
    \g_c[1+Y(\g_c)] = \g_c^{syn},
\end{equation}
have either a single solution or three different solutions. In most
physical scenarios Eq. \ref{EQ g_c general} has a single solution in
which case the standard definition of $\g_c$ holds, namely, $\g_c$
is the Lorentz factor above which electrons cool over the lifetime
of the system. In cases where Eq. \ref{EQ g_c general} has three
solutions, $\g_{c,min}<\g_{c,mid}<\g_{c,max}$, they satisfy
$\g_{c,min}<\g_{c,mid}<\g_0<\g_{c,max}=\g_c^{syn}$. Electrons with
$\g_{c,min}<\g<\g_{c,mid}$ cool fast by SSC emission, electrons with
$\g>\g_{c,max}$ cool fast by synchrotron emission and the rest of
the electrons (those with $\g<\g_{c,min}$ and those with
$\g_{c,mid}<\g<\g_{c,max}$) are not cooling over the system
lifetime. Below we present the solution of Eq. \ref{EQ g_c general}
for each of the cases covered in the previous sections, and discuss
the observational effects of cooling.

\subsection{Fast cooling}

{\it Case I ($\g_c<\g_m<\gmh$):} The value of
$Y(\g_c)=(\epse/\epsB)^{1/2}$ is independent of $\g_c$ and
therefore:
%\begin{equation}\label{EQ g_c1}
  %  \g_c \approx \frac{6 \pi m_e c}{\sigma_T B^2
  %  t}\sqrt{\frac{\epsB}{\epse}} .
%\end{equation}
\begin{equation}\label{EQ g_c1}
    \g_c \approx \g_c^{syn}
    \sqrt{\frac{\epsB}{\epse}} .
\end{equation}
is a solution of Eq. \ref{EQ g_c general}. This cooling frequency
always corresponds to SSC dominated cooling. In principle it is
possible to have three solutions to Eq. \ref{EQ g_c general} and
thus a second synchrotron dominated cooling frequency. However this
requires $\epse/\epsB>(\gmh/\g_m)^8(\g_m/\g_c)^{18}$ (where $\g_c$
is given by Eq. \ref{EQ g_c1}), which is unlikely to be satisfied in
astrophysical sources when $\gmh \gg \g_m$.

{\it Case IIa ($\gmh<\g_0<\g_{self}<\g_m$):} In this case, Eq.
\ref{EQ g_c general} always has a single solution and $\g_c$ is well
defined. If $\g_c^{syn}>\g_0$ then $\g_c=\g_c^{syn}$ and all the
electrons are cooling primarily by synchrotron and therefore SSC
cooling can be completely ignored. The synchrotron spectrum is then
as given in, e.g., \cite{Sari98}. If $\g_c^{syn}<\g_0$, electrons
that are cooling by SSC, i.e. with $\g<\g_0$, are upscattering
photons emitted by electrons that are cooling by synchrotron (i.e.
with Lorentz factor $\gt>\g_0$.) In summary:

\begin{equation}\label{EQ g_c2a}
    \g_c \approx \g_c^{syn} \left\{\begin{array}{cc}
    1 & \g_0<\g_c^{syn} \\
    \g_c^{syn}/\g_0 & (\gmh\g_0)^{1/2}<\g_c^{syn}<\g_0 \\
    \left(\reps\right)^{-1}  & \g_c^{syn}<(\gmh\g_0)^{1/2}
    \end{array} \right.
\end{equation}

If $\g_c$ is not small enough, it affects the Compton Y-parameter in
Eq. \ref{EQ Y2a} such that $Y(\g>\gch) \propto \g^{-4/3}$ and
affects the synchrotron spectrum in Eq. \ref{EQ Fsyn2a} such that
$F_\nu^{syn}(\nu<\nu_c) \propto \nu^{1/3}$.

{\it Case IIb ($\gmh<\g_{self}<\g_0<\g_m$):} In this case
Eq. \ref{EQ g_c general} may have more than one
solution. However, when there are three solutions only
$\g_{c,max}$ has an observable signature. The reason is that here
$\g_{c,max}<\g_m$ and therefore injected electrons (all with
$\g>\g_m$) do not have enough time to cool down below $\g_{c,max}$.
Thus, when there are three solutions to Eq. \ref{EQ g_c general} all
the electrons are cooling by synchrotron emission. The transition to
SSC cooling takes place when $Y(\g_{c,max}) \approx 1$ which is also
the point where $\g_{c,max} \approx \g_{c,mid}$ and there is
transition to a single solution of Eq. \ref{EQ g_c general} with
$\g_c \approx \g_{c,min}$. We denote the value of $\g_{c,max}$ and
$\g_{c,min}$ at this transition point as $\g_{c,max,0}$ and
$\g_{c,min,0}$. Since $\g_{c,min,0}$ can be much smaller than
$\g_{c,max,0}$ the observable effect of this transition is dramatic
where $\nu_c$ vary on a short time scale (comparable to $t$) by
orders of magnitude between $\nu_{syn}(\g_{c,max,0})$ and
$\nu_{syn}(\g_{c,min,0})$.

The value
$\g_{c,max,0}=(\epsilon_e/\epsilon_B)^{1/3}\g_m^{5/9}\gmh^{4/9}$ is
calculated using a synchrotron spectrum which is unaffected by SSC
cooling and the requirement $Y(\g_{c,max,0})
= 1$. (note that $\g_{c,max,0}$ is smaller than the value of $\g_0$
obtained assuming $\g_c < \gmh$). The value of $\g_{c,min,0}$ is
calculated using Eq. \ref{EQ Y2b} and the requirement $\g_{c,min,0}
Y(\g_{c,min,0})=\g_{c,max,0}$. If $\epse/\epsB<(\g_m/\gmh)^{5/6}$
then $\g_{c,min,0} = (\epse/\epsB)^{-4/3}\g_m^{10/9}\gmh^{-1/9}$,
otherwise $\g_{c,min,0} = (\epse/\epsB)^{-2/3}\g_m^{5/9}\gmh^{4/9}$.
The value of $\g_c$ in case IIb is therefore:
%\begin{equation}\label{EQ g_c2b}
  %  \g_c \approx \frac{6 \pi m_e c}{\sigma_T} \left\{\begin{array}{cc}
   % B^{-2} t^{-1} & \g_{c,max,0} < \g_c \\
    %\frac{6 \pi m_e c}{\sigma_T}\left(\reps\right)^{-2}\gmh^{-1} B^{-4} t^{-2} & \gmh<\g_c<\g_{c,min,0} \\
    %\left(\reps\right)^{-1}  B^{-2} t^{-1} & \g_c<\min\{\gmh,\g_{c,min,0}\}
   % \end{array} \right.
%\end{equation}
\begin{equation}\label{EQ g_c2b}
    \g_c \approx \g_c^{syn} \left\{\begin{array}{cc}
    1 & \g_{c,max,0} < \g_c \\
    \g_c^{syn} \gmh^{-1} \left(\reps\right)^{-2} & \gmh<\g_c<\g_{c,min,0} \\
    \left(\reps\right)^{-1} & \g_c<\min\{\gmh,\g_{c,min,0}\}
    \end{array} \right.
\end{equation}
where the second segment exist only if
$\epse/\epsB<(\g_m/\gmh)^{5/6}$, in which case $\gmh<\g_{c,min,0}$.
Note that $\g_c$ cannot obtain values between $\g_{c,max,0}$ and
$\g_{c,min,0}$.

The effect of $\g_c$ on the observed spectrum is as follow. When
$\g_c > \g_{c,max,0}$ all electrons are cooling by synchrotron and
the spectrum is therefore not affected at all by SSC and it is given
by e.g., \cite{Sari98}. When $\g_c < \g_{c,min,0}$ SSC cooling is
the dominated cooling process of low-energy electrons, and Eq.
\ref{EQ Y2b} can be used with the simple addition $Y(\g>\gch)
\propto \g^{-4/3}$. The power-law segment $F_\nu^{syn}(\nu<\nu_c)
\propto \nu^{1/3}$ should be added to Eq. \ref{EQ Fsyn2b}.

{\it Case IIc ($\gmh<\g_{self}<\g_m<\g_0$):} The effect of the
cooling frequency on the spectrum, and its behavior, are similar to
the previous case (IIb). In this case $\g_{c,max,0} =
(\epse/\epsB)^{3/7}\g_m^{3/7}\gmh^{4/7}$ if
$\epse/\epsB>(\g_m/\gmh)^{4/3}$ (note that $\g_{c,max,0}>\g_m$
here). Otherwise
$\g_{c,max,0}=(\epse/\epsB)^{1/3}\g_m^{5/9}\gmh^{4/9}$. We do not
give here the exact value of $\g_{c,min,0}$ but in all case
$\g_{c,min,0} < \gmh$. When $\g_c>\g_{c,max,0}$ all the electrons
are cooling by synchrotron and there is no SSC effect while $\g_c <
\g_{c,min,0}$ implies dominant SSC cooling of all electrons with
$\g<\g_0$ (including $\g_m$). $\g_c$ cannot assume a value between
$\g_{c,min,0}$ and $\g_{c,max,0}$ and the transition between
$\g_c=\g_{c,max,0}$ to $\g_c=\g_{c,min,0}$ is observed as a sudden
variation of $\nu_c$ by a few orders of magnitude on a short time
scale which also accompanied by significant variation in
$F_\nu(\nu_m)$ and therefore in the total synchrotron luminosity.
Note that if $\g_{c,max,0}>\g_m$ then this transition is also a
transition between slow and fast cooling regimes.

{\it Case III ($\gmh=\g_m$):} The behavior of the cooling frequency
is similar to the two previous cases with
$\g_{c,max,0}=\g_m(\epse/\epsB)^{3/7}$ and
$\g_{c,min,0}=\g_m(\epse/\epsB)^{-1/14}$ in case that $p>2.5$. If
$p<2.5$ then $\g_{c,max,0}$ is unchanged and
$\g_{c,min,0}=\g_m(\epse/\epsB)^{-3/14/(2p-2)}$. Similarly to
previous cases $\g_c$ can vary rapidly between $\g_{c,max,0}$ and
$\g_{c,min,0}$. Here this transition is also a transition between
slow and fast cooling regimes. In this case there is a regime where
there are three solutions to Eq. \ref{EQ g_c general} with
$\g_{c,min}<\g_m<\g_{c,mid}$. In such case there are electrons that
are cooling down to $\g_{c,min}$ and there are electrons with
$\g_{c,mid}<\g<\g_{c,max}$ that are not cooling. This regime is a
short transient (unless $B^2t$ is constant) and we do not present
here the resulting spectrum.

\subsection{Slow cooling}
We now describe how to find the value of $\g_c$ in the slow cooling
regime. In the following we assume $2<p<3$. If $\gmh\leqslant\g_m$
then the system is cooling slowly if $\g_c^{syn}$ is larger than
$\g_m$ in the cases IIa and IIb or larger than $\g_{c,max,0}$ in the
cases IIc and III. The value of $\g_c$ is then approximately
$\g_c^{syn}$ since $Y(\g_c)<1$.

If $\gmh \gg \g_m$ and $\epse/\epsB<\gmh/\g_m$  then there is always
a single solution to Eq. \ref{EQ g_c general}. If $\g_c^{syn}$ is
larger than $\gmh$ then $\g_c=\g_c^{syn}$. Otherwise Eq. \ref{EQ g_c
general} should be solved where the value of $Y(\g_c)$ (which may
still be smaller than 1) is evaluated using Eq. \ref{EQ. Yc slow}.

If $\gmh \gg \g_m$ but $\epse/\epsB>\gmh/\g_m$  then there may be
three solutions to Eq. \ref{EQ g_c general} during the slow cooling
phase. In this case
$\g_{c,max,0}=(\epse/\epsB)^{3/7}\g_m^{3/4}\gmh^{4/7}$ and
$\g_{c,min,0}$ is not much smaller than $\gmh$. Here, if
$\g_c^{syn}$ is larger than $\g_{c,max,0}$ then $\g_c=\g_c^{syn}$.
Otherwise Eq. \ref{EQ g_c general} should be solved where the value
of $Y(\g_c)$ is evaluated using Eq. \ref{EQ. Yc slow}. The result
then is smaller than $\gmh$. Observationally, similarly to previous
cases, $\g_c$ can vary on a short time scale between $\g_{c,max,0}$
and $\g_{c,min,0}$.

\section{Klein-Nishina effects in gamma-ray bursts}\label{SEC GRBs}
Synchrotron and/or SSC are most likely playing a major role in the
emission from both long and short GRBs \citep[for recent reviews see
][]{Piran05,Meszaros06,Nakar07}. GRB emission is composed of (at
least) two physically distinctive phases - the prompt and afterglow
emission. The prompt emission is observed as a short burst of
$\sim$MeV gamma-rays which lasts from a fraction of a second to
several minutes. The radiation process that generates these photons
is not determined yet  \cite[see recent discussion in][]{Piran08}
although a fast cooling synchrotron emission is a popular model. The
afterglow is observed for weeks and sometime even years in radio to
X-ray wavelength. It is most likely that the radio to X-ray
afterglow is a synchrotron emission (at first fast and later slow
cooling) from a relativistic shockwave that propagates into the
circum-burst medium.

\subsection{Prompt emission}
We consider here cases in which prompt emission of GRBs is generated
by fast cooling synchrotron emission. Since the observed spectra of
GRBs peak around a fraction of MeV we have approximately $h\nu_m
\sim m_e c^2$ implying that
\begin{equation}\label{EQ gmh GRB}
    \gmh \approx \Gamma .
\end{equation}
Opacity constraints \cite[e.g. ][]{Lithwick01} indicates that
$\Gamma \gtrsim 100$, while considerations such as the afterglow
onset suggest $\Gamma \lesssim 1000$. In the internal shock scenario
the radiating electrons are accelerated by mildly relativistic
shocks and therefore $\g_m \sim 100$. Thus, it is likely that in
some, and maybe even in a significant fraction, of the GRBs
$\gmh/\g_m \sim 1$ during the prompt emission and therefore, KN
effects play an important role in shaping the observed prompt MeV
spectrum. It is not clear however that the theory of optically thin
SSC emission that is discussed in this paper is directly applicable
to the prompt emission of all bursts. The reason is that here we
assume that the synchrotron photons are optically thin for pair
production of SSC photons. This assumption is justified if the
prompt emission is produced at rather large radii ($\sim
10^{15}$--$10^{16}$ cm), which is most likely in cases where the
variability timescale is longer than a few seconds. It is violated
however, in most cases, if the prompt emission is generated at
smaller radii of $\sim 10^{13}$ cm.

We therefore expect that our results are applicable at least in some
of the bursts. The expected observational KN signature may have
already been observed. The prompt emission spectrum of GRBs is
typically fitted by a broken power-law with a $F_\nu$ low-energy
spectral index $\alpha_L$ and high energy spectral index
$\alpha_{H}$. For long GRBs the value of $\alpha_L$ is between $-1$
and $1$ while the value of $\alpha_H$ is between $-0.5$  and $-2.5$
\citep{Preece00,Kaneko06}. The standard synchrotron fast cooling
model (that ignore KN effects) predicts $\alpha_L \approx -0.5$ and
$\alpha_H = -p/2$ which is clearly inconsistent with many of the
observed bursts \citep[e.g.,][]{Preece98,Kumar08}. \cite{Preece98}
point out that $\alpha_L > 1/3$ is inconsistent with optically thin
synchrotron spectrum (the so called ``line of death'') while $-0.5
<\alpha_L$ is inconsistent with synchrotron emission from fast
cooling electrons, which is required by the high efficiency of the
prompt phase. When KN effects are considered $\alpha_L$ can be as
high as $0$ and $\alpha_H$ can be as high as $-(p-1)/2$. Thus, KN
effects may be the reason for the hard spectrum observed in some of
the prompt emission spectra also in cases that we observe
synchrotron emission from fast cooling electrons, although it cannot
explain cases where $\alpha_L<0$.

\subsection{Afterglow}
Afterglow theories provide an approximate description of the
observed afterglow multi-wavelength light curve for a given set of
physical conditions (e.g., blast wave energy, external density
profile, initial jet structure, etc.). Incorporating the KN effects
described in sections \ref{SEC fastcooling}-\ref{SEC slowcooling} is
straight forward. Here we show the result of applying KN effects to
the external-shock theory of a spherical blast wave that propagates
into a constant external density \citep[e.g., ][]{Sari98}. Within
the framework of this theory:
\begin{equation}\label{EQ num_nuc_GRB}
\begin{array}{l}
    \nu_m= 5 \cdot 10^{11} \epsilon_{e,-1}^2 \epsilon_{B,-3}^{1/2}
    E_{53}^{1/2} T_{day}^{-3/2}~\rm Hz , \\
    \\
    \nu_c= 3 \cdot 10^{16} \epsilon_{B,-3}^{-3/2}
    E_{53}^{-1/2} n^{-1} T_{day}^{-1/2} [1+Y(\g_c)]^{-2}~\rm Hz ,
\end{array}
\end{equation}
where $E$ is the isotropic equivalent energy of the blast-wave, $n$
is the external medium density and $T_{day}$ is the observed time
since the explosion in units of days (we neglect here redshift
effects). $Q_{x}$ denotes the value of the quantity $Q$ in units of
$10^{x}$ (c.g.s). Note the explicit dependence of the r.h.s of the
equation for $\nu_c$ on $\g_c$.  The magnetic field in the shocked
fluid rest frame is $B\approx 0.1 \epsilon_{B,-3}^{1/2} E_{53}^{1/8}
n^{3/8} T_{day}^{-3/8}$ G, implying that
\begin{equation}\label{EQ g_self_afterglow}
    \g_{self} \approx 7.5 \cdot 10^5 \epsilon_{B,-3}^{1/6} E_{53}^{1/24} n^{1/8} T_{day}^{-1/8}
\end{equation}
depends very weakly on the physical parameters. Since electrons are
expected to be accelerated to much higher values than $\g_{self}$,
KN effects may affect the observed afterglow.

Starting several minutes after the
burst we expect the afterglow to be in its slow cooling phase.
During this phase we use Eqs. \ref{EQ gamma_hat} and \ref{EQ.
Yc slow} to find $Y(\g_c)$ and $\gch/\g_c$ in the various regimes:%these are approximated values. More accurate values can be found in SSC_KN3.tex
\begin{equation}\label{EQ Yc_GRB}
    Y(\g_c) \approx \left\{
    \begin{array}{lr}
      5\;e^\frac{4(2.4-p)}{4-p}\;\epsilon_{e,-1}^\frac{p-1}{4-p}\;\epsilon_{B,-3}^{-\frac{3-p}{4-p}}\;
      E_{53}^\frac{p-2}{2(4-p)}\;n^\frac{p-2}{2(4-p)}\;T_{day}^{-\frac{p-2}{2(4-p)}} & ~\gch\gg\g_c ~;~ Y(\g_c)\gg1  \\

      e^\frac{27.5-9.7p}{p-1}\;\epsilon_{e,-1}^2\;\epsilon_{B,-3}^{\frac{3-p}{2(p-1)}}\;
      E_{53}^\frac{1}{p-1}\;n^\frac{5-p}{2(p-1)}\;T_{day}^{-\frac{p-2}{p-1}} & ~\gch\ll\g_c ~;~ Y(\g_c)\gg1\\

      e^{15.4-5.4p}\;\epsilon_{e,-1}^{p-1}\;\epsilon_{B,-3}^{-(3-p)}\;E_{53}^{\frac{p-2}{2}}\;
      n^\frac{p-2}{2}\;T_{day}^{-\frac{p-2}{2}} & ~\gch\gg\g_c ~;~ Y(\g_c)\ll1\\

      e^{13.8-4.8p}\;\epsilon_{e,-1}^{p-1}\;\epsilon_{B,-3}^{\frac{3-p}{4}}\;
      E_{53}^\frac{1}{2}\;n^\frac{5-p}{4}\;T_{day}^{-\frac{p-2}{2}} & ~\gch\ll\g_c ~;~ Y(\g_c)\ll1
    \end{array}\right.
\end{equation}
and
\begin{equation}\label{EQ gch_GRB}
    \frac{\gch}{\g_c} \approx \left\{
    \begin{array}{lr}
      35\;e^\frac{11(2.4-p)}{4-p}\;\epsilon_{e,-1}^{\frac{3(p-1)}{4-p}}\;\epsilon_{B,-3}^{\frac{p+2}{2(4-p)}}
      \;E_{53}^\frac{p+2}{2(4-p)}\;n^{\frac{3}{(4-p)}}\;T_{day}^{-\frac{3(p-2)}{2(4-p)}} & ~\gch\gg\g_c ~;~ Y(\g_c)>1  \\

      0.35\;e^\frac{83-29p}{p-1}\;\epsilon_{e,-1}^6\;\epsilon_{B,-3}^{\frac{p+2}{p-1}}\;
      E_{53}^\frac{p+2}{p-1}\;n^\frac{6}{p-1}\;T_{day}^{-\frac{3(p-2)}{p-1}} & ~\gch\ll\g_c ~;~ Y(\g_c)>1\\

      0.35 \;\epsilon_{B,-3}^{5/2}\; E_{53}\; n^{3/2}& Y(\g_c)<1
    \end{array}\right. .
\end{equation}
For typical sets of long GRB parameters ($\epse \sim 0.1$, $\epsB
\sim 10^{-3}$--$10^{-2}$, $p=2.1$--$2.7$, $E_{53}=0.1$--$10$, and $n=0.1$--$1
{\rm cm^{-3}}$) we have $Y(\g_c) \sim 1$--$10$ and $\gch/\g_c \gtrsim
1$. This implies that the SSC energy output of bright long GRBs
(assuming canonical physical parameters) is only weakly affected by
the KN limit. The cooling frequency in this regime ($Y(\g_c)> 1$ and
$\gch>\g_c$) is:
\begin{equation}\label{EQ KNnu_c GRB}
    \nu_c \approx 10^{15}\;e^{\frac{8(p-2.4)}{4-p}}\;\epsilon_{e,-1}^{-\frac{2(p-1)}{4-p}}\;
    \epsilon_{B,-3}^{-\frac{p}{2(4-p)}}\;E_{53}^{-\frac{p}{2(4-p)}}\;n^{-\frac{2}{4-p}}\;
    T_{day}^{-\frac{(8-3p)}{2(4-p)}} {\rm Hz} ,
\end{equation}
which for $p \neq 2$ can be significantly different than the
standard expression when SSC cooling is included and KN effects are
neglected. For example here $\nu_c$ is independent of $T$ for
$p=8/3$ where in the standard model $\nu_c \propto T^{-1/2}$ for any
$p$ value. The synchrotron spectrum is affected at
\begin{equation}\label{EQ KNnu_ch GRB}
    \nuch \approx 10^{18}\;e^{\frac{15(2.4-p)}{4-p}}\;\epsilon_{e,-1}^{\frac{4(p-1)}{4-p}}\;
    \epsilon_{B,-3}^{\frac{4+p}{2(4-p)}}\;E_{53}^{\frac{4+p}{2(4-p)}}\;n^{\frac{4}{4-p}}\;
    T_{day}^{-\frac{(3p-4)}{2(4-p)}} {\rm Hz},
\end{equation}
where the synchrotron spectral index becomes slightly harder
($-0.75(p-1)$ instead of $-p/2$). This KN signature of spectral
hardening may be observed in the X-ray.

$\epsB$ is one of the least constrained parameters in GRB external
shocks and while $\epsB = 10^{-3}-10^{-2}$ seems to be a favorable
value, $\epsB$ may be significantly smaller. However, smaller
$\epsB$ does not necessarily imply larger value of $Y(\g_c)$,
whereas, naively, the opposite is expected by a model that neglects
KN effects where $Y(\g)=Y_{noKN}=\sqrt{\epse/\epsB}$. Eq. \ref{EQ
Yc_GRB} shows that for $\gch>\g_c$ the value of $Y(\g_c)$ increases
when $\epsB$ is reduced (assuming $Y(\g_c)>1$). However, for
$\gch<\g_c$ the value of $Y(\g_c)$ decreases when $\epsB$ is
reduced. The reason is that in the latter regime the KN suppression
becomes stronger when $\epsB$ is reduced. As a result, if we allow
$\epsB$ to vary, while holding the rest of the parameters constant,
$Y(\g_c)$ is maximized once $\gch=\g_c$:
\begin{equation}\label{EQ Ymax_GRB}
   Y_{max}(\g_c)\approx 10\;e^\frac{27(3.2-p)^2-17}{(p+2)(4-p)}\;\epsilon_{e,-1}^{\frac{5(p-1)}{p+2}}\;
   E_{53}^{1/2}\;n^\frac{8-p}{2(p+2)}\;T_{day}^{-\frac{5(p-2)}{2(p+2)}} .
\end{equation}
For the canonical values of the parameters taken here ($E_{53}=1$,
$\epse=0.1$ and $n=1 {\rm cm^{-3}}$) this maximum is achieved at
$\epsB \sim 10^{-4}$. Eq. \ref{EQ Ymax_GRB} implies that in bursts
that are not very bright ($E<10^{51}$ erg and $n<1 {\rm cm^{-3}}$ )
the SSC cooling is suppressed for any value of $\epsB$ and the
synchrotron emission is better approximated by ignoring SSC cooling
than by including it but ignoring KN effects. It also implies that
there is no significant SSC GeV emission accompanying late
afterglows that are not very luminous in radio to X-ray emission.
This includes most (and maybe all) of the short GRBs \citep[see also
discussion in][]{Nakar07}.

During the first several minutes the afterglow is often cooling
fast, in which case the importance of the KN effects are determined
by the ratio $\gmh/\g_m$:
\begin{equation}\label{EQ gmh_GRB}
    \frac{\gmh}{\g_m} \approx  100\;\epsilon_{e,-1}^{-3}\;\epsilon_{B,-3}^{-1/2}
    \;E_{53}^{-1/2}\;T_{2}^{3/2},
\end{equation}
where $T_2=T/100$ s. This implies that according to the standard
external shock model the energy output of the SSC emission during
the early afterglow is not strongly affected by the KN limit.

\section{Summary}\label{SEC summary}
In this paper we present analytic approximation to the optically
thin synchrotron-SSC spectra in case that the distribution of the
radiating electrons is modified by the KN cross-section. We consider
here cases where there is only a single SSC scattering and
multiple photon upscatterings are entirely suppressed by the KN
cross-section. We also consider only systems which are optically
thin to pair production by the SSC photons. These analytic
expressions are useful for construction of analytic and semianalytic
theory of radiation from astrophysical sources, such as GRBs, AGNs
and pulsar wind nebula, where KN effects may be important.

We find six general spectral types (with some subcases within these
six types) that differ mostly by the level of SSC suppression by the
KN cross-section. Table \ref{Table physical cond} summarizes the
physical conditions that result in each of the spectral types. It
also points to the relevant section in the paper that discusses the
case as well as the relevant equations of $Y(\g)$, $F_\nu^{syn}$ and
$F_\nu^{IC}$.

The main effect of the KN limit on the electron distribution is the
additional dependence of the cooling rate on the electron Lorentz
factor. Electrons with higher Lorentz factor can upscatter a smaller
fraction of the synchrotron photons and are therefore cooling more
slowly compared to the case where the KN limit is ignored. The
result is the introduction of new critical values of Lorentz factors
which correspond to new synchrotron and SSC power-law segments. The
new critical Lorentz factors differ between the spectral cases and
are typically one or more of $\numh$, $\nuch$, $\nu_0$ and $\nuoh$,
($\g_0$ is defined so $Y(\g_0)=1$ and $\gh$ is defined in Eq.
\ref{EQ gamma_hat}). Table \ref{Table Critical gammas} summarizes
the value of the critical Lorentz factors in each of the cases as a
function of $\epse/\epsB$, $\g_m$, $\g_c$, $\gmh$ and $\gch$. The
new power-law indices, and there frequency range, that are
introduced to the synchrotron spectrum by the KN limit are
summarized in table \ref{Table synch PLSs}. The ratio of inverse
Compton to synchrotron cooling, i.e. $Y(\gamma)$, is always a
decreasing function. Therefore, synchrotron power-law segments that
are introduced by KN effects are always harder than, or comparable
to, those that are predicted by standard synchrotron theories, which
ignore the KN limit.

The hardening of the synchrotron spectrum results in two main
observable features that should be taken into account when an
observed synchrotron spectrum is analyzed. First, a spectrum of the
fast cooling regime can asymptotically be as hard as $F_\nu \propto
\nu^0$. Such a power-law segment is observed when the electrons
which are radiating the synchrotron photons within this power-law
segment are cooled down predominantly by upscattering synchrotron
photons within this power-law segment itself (cases IIb and IIc).
This flat spectrum is significantly different than any of the
predictions of a standard synchrotron theory that ignore KN effects,
as standard theory predicts spectral indices that are harder than
$1/3$ or softer than $-1/2$. Note that the actual spectrum in this
specific segment is significantly smoother than the analytic
approximation. As a result $\alpha$ in this power-law segment may
approach the asymptotic value $\alpha=0$ only for very large values
of $\epse/\epsB$. For example, numerical solution of equations
\ref{EQ cool func}, \ref{EQ dN_dt} and \ref{EQ Ynumeric} shows that
if $\epse/\epsB=100$ the $\alpha$ value in this segment cannot
exceed $\approx -0.3$ while if $\epse/\epsB=10^4$ it cannot exceed
$\approx -0.2$. Second, $\nu_{syn}^{peak}$ is typically an available
observable even when the spectral resolution is limited, and
standard theory associates it with $\nu_m$ in the fast cooling
regime and $\nu_c$ in the slow cooling regime. KN effects can modify
this interpretation as the suppressed cooling of higher energy
electrons may results in $\nu_{syn}^{peak} \approx \nu_0 >
\max\{\nu_m,\nu_c\}$.

The SSC spectrum is also modified by KN limit. The most observable
signature of KN effects in the SSC spectrum is the ``KN cut-off,''
which is actually not a sharp cut-off but a consecutive set of
power-law segments that become steeper at higher frequencies. The
power-law segments and their physical origin are summarized in table
\ref{Table SSC PLSs}. In all cases the SSC energy output is
dominated by $\g_m$ [$\g_c$] electrons in the fast [slow] cooling
regime. Therefore emission at $\nu>\nu^{IC}_{peak}$ that is affected
by the KN limit is dominated by electrons with $\g>\g_m$[$\g>\g_c$]
in the fast[slow] cooling regime. If these electrons are still
cooling predominantly by SSC emission (i.e., their $\g<\g_0$) then
the first KN break at $\nu>\nu^{IC}_{peak}$ is to a very mild
spectral index, $-p+1$, and if $p \approx 2$ it is hardly
distinguishable from the $-p/2$ spectral index expected in case that
KN effects are unimportant. A clear steepening in the
light curve can be observed only once the Lorentz factor of the
upscattering electron is $\g>\gmh$ [$\g>\gch$] in the fast [slow]
cooling regime. At this point the SSC spectral index depends on the
synchrotron spectral index of the upscttered photons (at $\nuh$) and
it ranges between $-p+1/2$ to $-p-1/3$. The value of
$\nu^{IC}_{peak}$ is also affected by the KN cross section when
$\gmh<\g_m$ [$\gch<\g_c$] in the fast [slow] cooling regime. In these
cases $\nu^{IC}_{peak} \approx 2\nu_m\g_m\gmh$ [$\nu^{IC}_{peak}
\approx 2\nu_c\g_c\gch$] instead of the standard value ($\approx
2\nu_m\g_m^2$ [$\approx 2\nu_c\g_c^2$]).

The KN cross-section can also affect the value of $\nu_c$ and in
some cases can result in  a unique temporal evolution---$\nu_c$ can
``jump" by orders of magnitude over a short time scale. When KN
effects are ignored the cooling rate of an electron always increases
with its energy.  The KN modified SSC cooling can revise this
property introducing similar cooling rates for high-energy electrons
that cool by synchrotron emission and much lower energy electrons
that cool by SSC emission, where the cooling rate of intermediate
energy electrons is much lower. When the similar cooling time of the
high-energy electrons and low-energy electrons in such configuration
becomes comparable to the system lifetime, $t$ there is a sudden
change of the observed $\nu_c$ between the frequency that
corresponds to high-energy synchrotron cooling electrons
($\g_{c,max,0}$) and the frequency that corresponds to low-energy
SSC cooling electrons ($\g_{c,min,0}$). This transition between
$\g_{c,max,0}$ and $\g_{c,min,0}$ is completed over a short time
scale (comparable to $t$). Table \ref{Table Critical gammas}
summarizes the values of $\g_{c,max,0}$ and $\g_{c,min,0}$ for cases
in which such sudden variation may be observed.

We demonstrate an application of our results to the a synchrotron
model of the prompt emission of GRBs. We find that signature of KN
effects may have already been observed in the prompt emission of
some GRBs, in the form of spectrum that is harder than the standard
synchrotron model both above and below $\nu_{syn}^{peak}$. We
examine also the application of the KN limit to the external shock
model (in its quasi-spherical regime) of GRB afterglow emission.
Here we find that in the slow cooling phase,  assuming canonical
parameters: (a) $\nuch$ may be observed passing through the X-ray
(b) the value of $\nu_c$ may be significantly modified by the KN
limit (c) the SSC GeV energy output is unlikely to be suppressed by
the KN cross-section when the radio-X-ray afterglow is luminous, but
it may be strongly suppressed when the radio-X-ray afterglow is
faint. Finally we find that this afterglow model does not predict a
large SSC-to-synchrotron luminosity ratio even in case that the
poorly constrained model parameter $\epsB$ is very small. The reason
is that a low value of $\epsB$ results in a strong KN suppression of
the SSC luminosity. This result implies that $\epsB$ cannot be
easily deduced even if the SSC to synchrotron luminosity ratio of
slow cooling GRB afterglow is accurately measured.

%acknowledgement
We thank Tsvi Piran, Pawan Kumar  and Orly Gnat for helpful
discussions. R.S. was partially supported by ERC, ATP and IRG
grants, and a Packard Fellowships. This research was supported by
the Israel Science Foundation (grant No. 174/08).

\newpage
\begin{table*}[h]
\begin{footnotesize}
%\begin{minipage}{70mm}
\caption{} \label{Table physical cond}
\begin{tabular}{@{}|c|c|c|c|c|c|c|c|c|@{}}
\hline
Case & \mcol{2}{c|}{\cent Cond 1} & $^{\dagger\dagger}$Cond 2 & Cond 3 & \S & $^\dagger$$Y$ & $^\dagger$$F_\nu^{syn}$ & $^\dagger$$F_\nu^{IC}$ \\
\hline\hline
\mrow{5}{10mm}{\cent Case I (Fast)} & \mrow{5}{14mm}{\cent $\g_m < \gmh$} & $\sqrt{\reps}<\frac{\g_m}{\g_c}$ & \multirow{5}{*}{\cent $\g_c^{syn}<\sqrt{\reps}\g_m$} & \multirow{3}{*}{\cent $\left(\frac{\g_0}{\g_m}\right)^{2p-4}>\reps$} & \ref{SEC caseI} &  \ref{EQ Y1} & \ref{EQ Fsyn1a} &  \ref{EQ FIC1a} \\
\cline{3-3}\cline{6-9}%
&  & $\sqrt{\reps}>\frac{\g_m}{\g_c}$ & & &\ref{SEC caseI} & \ref{EQ Y1} &\ref{EQ Fsyn1b} & \ref{EQ FIC1b}  \\
\cline{3-3}\cline{5-9}%
&  & $\sqrt{\reps}<\frac{\g_m}{\g_c}$ & & $\left(\frac{\g_0}{\g_m}\right)^{2p-4}<\reps$ & \ref{SEC caseIp=2} & \ref{EQ Y1}+\ref{EQ YF1p=2} & \ref{EQ Fsyn1a}+\ref{EQ YF1p=2} & -  \\
\hline
\mrow{2}{14mm}{\cent Case IIa (Fast)}&\mcol{2}{c|}{\cent \mrow{2}{*}{\cent $\frac{\epse}{\epsB}<\left(\frac{\g_m}{\gmh}\right)^\frac{1}{3}$ }}& \mrow{2}{*}{\cent $\g_c^{syn}<\g_m$} & &\mrow{2}{*}{\cent\ref{SEC caseIIa}} & \mrow{2}{*}{\cent\ref{EQ Y2a}} &\mrow{2}{*}{\cent\ref{EQ Fsyn2a}} & \mrow{2}{*}{\cent text} \\
&\mcol{2}{c|}{}&&&&&& \\
\hline
\mrow{2}{14mm}{\cent Case IIb (Fast)}&\mcol{2}{c|}{\cent \mrow{2}{*}{\cent $\left(\frac{\g_m}{\gmh}\right)^\frac{1}{3}<\frac{\epse}{\epsB}<\frac{\g_m}{\gmh}$ }}& \mrow{2}{*}{\cent $\g_c^{syn}<\g_m$} & &\mrow{2}{*}{\cent\ref{SEC caseIIa}} & \mrow{2}{*}{\cent\ref{EQ Y2b}} &\mrow{2}{*}{\cent\ref{EQ Fsyn2b}} & \mrow{2}{*}{\cent text}\\
&\mcol{2}{c|}{}&&&&&& \\
\hline
\mrow{2}{14mm}{\cent Case IIc (Fast)}&\mcol{2}{c|}{\cent  \mrow{2}{*}{\cent $\frac{\g_m}{\gmh}<\frac{\epse}{\epsB}<\left(\frac{\g_m}{\gmh}\right)^3$ }}& \mrow{2}{*}{\cent $\g_c^{syn}<\g_{c,max,0}$} & $2<p<3$ &\mrow{2}{*}{\cent \ref{SEC caseIIc}} & \ref{EQ Y2c} &\ref{EQ Fsyn2c} & text \\
\cline{5-5}\cline{7-9}%
&\mcol{2}{c|}{}&&$p>3$&&text&text&- \\
\hline
\mrow{2}{14mm}{\cent Case III (Fast)}& \mcol{2}{c|}{\cent \mrow{2}{*}{$\g_m=\gmh$}} & \mrow{2}{*}{\cent $\g_c^{syn}<\g_{c,max,0}$} & $2<p<2.5$ &\mrow{2}{*}{\cent \ref{SEC caseIII}} & \ref{EQ Y3} &\ref{EQ Fsyn3} & \ref{EQ FIC3} \\
\cline{5-5}\cline{7-9}%
&\mcol{2}{c|}{}&&$p>2.5$&&text&text&text \\
\hline
\mrow{2}{14mm}{\cent Slow Cooling}& \mcol{2}{c|}{} & \mrow{2}{*}{\cent Not fast} &  \mrow{2}{*}{\cent $2<p<3$}  &\mrow{2}{*}{\cent \ref{SEC slowcooling}} & \mrow{2}{*}{\cent \ref{EQ. Yslow}+\ref{EQ. Yc slow}} &\mrow{2}{*}{\cent \ref{EQ FsynSlow1}} &\mrow{2}{*}{\cent \ref{EQ FIC slow}} \\
&\mcol{2}{c|}{}&&&&&&\\
\hline
\end{tabular}
\end{footnotesize}
\newline\newline
A summary of the physical conditions that correspond to any of the
spectral cases and sub-cases discussed in the paper. In order to
find the relevant spectrum the conditions are applied from left to
right in the following logic sequence: {\it ``if cond 1, then if
cond 2, then if cond 3, then ..."}. If all conditions are satisfied
then the relevant case is discussed in section ``\S"  and the
relevant equation numbers of $Y$, $F_\nu^{syn}$
and $F_\nu^{IC}$ are indicated.\\
$^\dagger$ The number of the relevant equation. When the entree is
{\it ``text"} then this case is discussed in the text of the
relevant
\S. \\
$^{\dagger\dagger}$ The condition for fast or slow cooling regime.
$\g_c^{syn} \equiv (6 \pi m_e c)/(\sigma_T B^2 t)$ depends on the
physical parameters of the system (it is the value of the
synchrotron cooling frequency if SSC emission is ignored). The
spectrum is in the slow cooling regime in case that {\it cond 1} of
any of the fast cooling phases is satisfied but the corresponding
{\it cond 2} is not satisfied (e.g., if $\g_m<\gmh$ but
$\g_c^{syn}>\g_m\sqrt{\epse/\epsB}$).

%\end{minipage}
\end{table*}

\begin{table*}
\begin{footnotesize}
% \begin{minipage}{70mm}
\caption{Critical Lorentz factor values} \label{Table Critical gammas}
\begin{tabular}{@{}|c|c|c|c|c|c|@{}}
\hline
Case &  & $\g_0$ & $\goh$ & $\g_{c,max,0}$ & $\g_{c,min,0}$\\
\hline\hline
\mrow{3}{10mm}{\cent Case I (Fast)} & $\sqrt{\reps}<\frac{\g_m}{\g_c}$ & $\gmh\reps$ & $^\dagger~\frac{\g_m^2}{\gmh}\left(\reps\right)^{-2}$& - & -  \\
\cline{2-6}%
&$\sqrt{\reps}>\frac{\g_m}{\g_c}$ & $\gch\left(\frac{\epse\g_c^2}{\epsB\g_m^2}\right)^\frac{3}{8}$& $^\dagger~\frac{\g_m^{3/2}\g_c^{1/2}}{\gch}\left(\reps\right)^{-\frac{3}{4}}$  & - & - \\
\hline
\mrow{2}{14mm}{\cent Case IIa (Fast)}& &\mrow{2}{*}{\cent $\gmh\left(\reps\right)^2$} & & -& -\\
&&&&& \\
\hline
\mrow{3}{14mm}{\cent Case IIb (Fast)}& $\reps<\left(\frac{\g_m}{\gmh}\right)^\frac{5}{6}$& \mrow{3}{*}{\cent $\sqrt{\reps\g_m\gmh}$} &\mrow{3}{*}{\cent $\g_m\left(\reps\right)^{-1}$}& \mrow{3}{*}{\cent $\left(\reps\right)^{\frac{1}{3}}\g_m^{5/9}\gmh^{4/9}$} &{\scriptsize $\left(\reps\right)^{-\frac{4}{3}}\g_m^\frac{10}{9}\gmh^{-\frac{1}{9}}$}\\
\cline{2-2}\cline{6-6}%
&$\reps>\left(\frac{\g_m}{\gmh}\right)^\frac{5}{6}$&&&&$\left(\reps\right)^{-\frac{2}{3}}\g_m^{5/9}\gmh^{4/9}$\\
\hline
\mrow{2}{14mm}{\cent Case IIc (Fast)}&$\reps<\left(\frac{\g_m}{\gmh}\right)^\frac{4}{3}$& \mrow{3}{*}{\cent $\sqrt{\reps\g_m\gmh}$} &\mrow{3}{*}{\cent $\g_m\left(\reps\right)^{-1}$}& $\left(\reps\right)^{\frac{1}{3}}\g_m^{5/9}\gmh^{4/9}$ & \mrow{3}{*}{\cent $<\gmh$}\\
\cline{2-2}\cline{5-5}%
&$\reps>\left(\frac{\g_m}{\gmh}\right)^\frac{4}{3}$&&& $\left(\reps\right)^{\frac{3}{7}}\g_m^{3/7}\gmh^{4/7}$ &\\
\hline
\mrow{2}{14mm}{\cent Case III (Fast)}& $p<2.5$ & $\g_m\left(\reps\right)^{\frac{3}{2(4-p)}}$ & $\g_m\left(\reps\right)^{-\frac{3}{4-p}}$ & $\g_m \left(\reps\right)^{3/7}$&$\left(\reps\right)^{-\frac{3}{14(2p-2)}}$\\
\cline{2-6}%
& $p>2.5$ & $\g_m \reps$& & $\g_m \left(\reps\right)^{3/7}$&$\left(\reps\right)^{-1/14}$\\
\hline
\mrow{2}{14mm}{\cent Slow Cooling}& \mrow{2}{*}{\cent $2<p<3$} & \mrow{2}{*}{\cent text}& \mrow{2}{*}{\cent text} & \mrow{2}{*}{\cent text} &\mrow{2}{*}{\cent text}\\
&&&&&\\
\hline
\end{tabular}
\end{footnotesize}
\newline\newline
Critical values of $\g$ that have an observable signature. These are
given as a function of the ratio $\epse/\epsB$ and critical Lorentz
factors $\g_m$, $\gmh$, $\g_c$ and $\gch$. The observed signature of
$\g_0$ and $\goh$ (when indicated) is a spectral break in $\nu_0$
and $\nuoh$. The observed value of $\g_c$ is always larger than
$\g_{c,max,0}$ or smaller than $\g_{c,min,0}$ as intermediate value
are not observed (see \S\ref{SEC g_c}).
%\end{minipage}
\end{table*}

\begin{table*}
\begin{footnotesize}
% \begin{minipage}{70mm}
\caption{Synchrotron Power-law indices that result from KN effects}
\label{Table synch PLSs}
\begin{tabular}{@{}|c|c|c|@{}}
\hline
$^a$Index & $^b$Case - frequency range & $^c$Origin \\
\hline\hline
\mrow{2}{*}{\cent $0$}& IIb - $\nuoh<\nu<\nu_0$& \mrow{2}{*}{\cent $\g_c<\g<\g_m$ electrons cool on photons from their own PLS}\\
&IIc - $\numh<\nu<\nu_m$& \\
\hline
\mrow{2}{*}{\cent $-\frac{1}{4}$}& IIa - $\numh<\nu<\nu_0$&\mrow{2}{*}{\cent $\g_c<\g<\g_m$ electrons cool on $F_\nu \propto \nu^{1/2}$ photons}\\
&IIb - $\numh<\nu<\nuoh$&\\
\hline
$-\frac{p-1}{4}$& IIc - $\nuoh<\nu<\numh$& $\g_c<\g<\g_m$ electrons cool on $F_\nu \propto \nu^{(p-1)/2}$ photons\\
\hline
$-\frac{p-1}{3}$& III ($p<2.5$) - $\nuoh<\nu<\nu_m$& $\g_c<\g<\g_m$ electrons cool on $F_\nu \propto \nu^{2(p-1)/3}$ photons \\
\hline
\mrow{2}{*}{\cent$-\frac{p}{2}+\frac{2}{3}$}& I - $\nuch<\nu<\nu_0$&\mrow{2}{*}{\cent $\max\{\g_c,\g_m\}<\g$ electrons cool on $\nu<\min\{\nu_c,\nu_m\}$ photons}\\
& Slow - $\numh<\nu<\nu_0$&\\
\hline
$-\frac{p}{2}+\frac{1}{2}$& IIc - $\nu_m<\nu<\nu_0$& $\max\{\g_c,\g_m\}<\g$ electrons cool on $F_\nu \propto \nu^{0}$ photons\\
\hline
$-\frac{2(p-1)}{3}$& III - $\nu_m<\nu<\nu_0$&$\max\{\g_c,\g_m\}<\g$ electrons cool on $F_\nu \propto \nu^{(p-1)/3}$ photons\\
\hline
$-\frac{p}{2}+\frac{1}{4}$& I - $\numh<\nu<\nuch$&$\max\{\g_c,\g_m\}<\g$ electrons cool on $F_\nu \propto \nu^{1/2}$ photons\\
\hline
\mrow{2}{*}{\cent $-\frac{3(p-1)}{4}$}& Slow cooling -& \mrow{2}{*}{\cent $\max\{\g_c,\g_m\}<\g$ electrons cool on $F_\nu \propto \nu^{(p-1)/2}$ photons}\\
&$\max\{\nuch,\nu_c\}<\nu<\min\{\numh,\nu_0\}$&\\
\hline
\end{tabular}
\end{footnotesize}
\newline
%\end{minipage}
The power-law segments (PLSs) that are introduced to the synchrotron
spectrum by the KN limit. \\
$^a$ The power-law index of $F_\nu$\\
$^b$ The case and frequency range in which this PLS is observed.\\
$^c$ The physical origin of the corresponding PLS, i.e., the range
of electron Lorentz factor values that dominate the synchrotron
emission and the PLS of the synchrotron photons that dominates the
cooling of these electrons.
\end{table*}

\begin{table*}
\begin{footnotesize}
%\begin{minipage}{140mm}
\caption{SSC power-law indices at $\nu>\nu_{peak}^{IC}$}
\label{Table SSC PLSs}
\begin{tabular}{@{}|c|c|@{}}
\hline
Index & Origin (all upscattering electrons have $\max\{\g_c,\g_m\}<\g$) \\
\hline\hline
\mrow{2}{*}{\cent $-p+1$} & $\gmh<\g<\g_0$ (fast cool)\\
&$\gch<\g<\g_0$ (slow cool)\\
\hline
$-p+\frac{1}{2}$& upscattering photons with $F_\nu^{syn} \propto \nu^{-1/2}$\\
\hline
$-p+\frac{1}{4}$& upscattering photons with $F_\nu^{syn} \propto \nu^{-1/4}$\\
\hline
$-p$& upscattering photons with $F_\nu^{syn} \propto \nu^{0}$\\
\hline
$-p-\frac{1}{3}$& upscattering photons with $F_\nu^{syn} \propto \nu^{1/3}$\\
\hline
\end{tabular}
\end{footnotesize}
\newline\newline
The SSC power-law indices of $F_\nu^{IC}$  at $\nu>\nu_{peak}^{IC}$,
which are a result of the KN limit. In all cases the electrons that
dominate the emission in this range have lorentz factors
$\g>\max\{\g_c,\g_m\}$. The  power-law segment in the first row,
$F_\nu^{IC} \propto \nu^{-p+1}$, is observed whenever the
upscattering electrons are within the Lorentz factor range indicated
in the second columns. The other four power-law segments are
observed whenever the SSC luminosity of electrons is dominated by
upscattering photons from the synchrotron power-law indicated in the
second column. See text for more details.
%\end{minipage}

\end{table*}

\newpage
%\bibliographystyle{apj}
%\bibliography{SSC_KN}

\end{document}